\documentclass[aps,pra,showpacs,twocolumn,eqsecnum]{revtex4}
\usepackage[dvips,final]{graphicx}
\usepackage{bm, graphics, amsmath}
\begin{document}

\title{Optimal measurements for the discrimination of quantum states with a fixed rate of inconclusive results}

\author{Ulrike Herzog}
\affiliation {Nano-Optics, Institut f\"ur Physik,
Humboldt-Universit\"at Berlin, Newtonstrasse 15, D-12489 Berlin,
Germany}


\begin{abstract}
We study the discrimination of $N$ mixed quantum states in an optimal measurement that  maximizes the probability of correct  results while the probability of inconclusive results is fixed at a given value.  After considering  the discrimination of $N$ states in a $d$-dimensional Hilbert space, we focus on the discrimination of qubit states. We develop a method to determine an optimal measurement for discriminating arbitrary qubit states, taking into account that often the optimal measurement is not unique and the maximum probability of correct results can be achieved by several different measurements.  Analytical results are derived for a number of examples, mostly for the discrimination  between qubit states which possess a partial symmetry, but also for discriminating $N$ equiprobable qubit states  and for the dicrimination between a pure and a uniformly mixed state in $d$ dimensions.    
In the special case where the fixed rate of inconclusive results is equal to zero, our method provides a treatment for the minimum-error discrimination of arbitrary qubit states which differs from previous approaches.
\end{abstract}


\maketitle

\section{Introduction }
The discrimination between a finite number of quantum states, prepared with given prior probabilities, is essential for many tasks in quantum information and quantum cryptography.
Since nonorthogonal states cannot be distinguished perfectly, discrimination strategies have been developed which  are  optimal with respect to various figures of merit. Based on the outcome of a measurement, in these strategies a guess is made about the actual state of the quantum system. Several strategies admit a certain probability, or rate,  $Q$, of inconclusive results, where the measurement outcomes do not allow to infer the state.

In the strategy of minimum-error discrimination \cite{helstrom,holevo} 
inconclusive results are not permitted, and the overall  probability $P_e$ of making a wrong guess is minimized with $Q=0$, which corresponds to maximizing the overall probability $P_c$ of getting a correct result. Apart from studying some general features of the optimal measurement \cite{hunter1,hunter}, in the beginning mainly the minimum-error discrimination of states obeying certain symmetry properties or of  two mixed states was investigated, see e. g. \cite{ban, barnett01, herzog-bergou, andersson,  chou, eldar, herzog0, herzog-bergou1}. The minimum-error discrimination of more than two states that are arbitrary  has gained renewed interest only recently \cite{samsonov, deconinck, kimura, jafarizadeh,  bae1,ha,ha1}. 

 Generalizing the concept of minimum-error discrimination,  an optimal discrimination strategy has been studied which maximizes the total rate of correct results, $P_c$, with a  fixed  rate $Q$ of inconclusive results \cite{chefles-barnett,  zhang-li, fiurasek, eldar1, herzog3, bagan, nakahira}.
This strategy also maximizes the relative rate  $P_c/(1-Q)$ for the fixed value $Q$,  that is it yields the maximum achievable fraction of correct results referred to all conclusive results, which in most cases is larger for $Q>0$ than for $Q=0.$   
Necessary and sufficient operator conditions for the optimal measurement have been derived \cite{fiurasek, eldar1, herzog3}, and it was shown  \cite{herzog3, bagan, herzog2} that  under certain conditions the strategy for maximizing $P_c$ with a fixed rate $Q$  contains the optimal strategies for unambiguous discrimination \cite{ivan,dieks,peres}  or for discrimination with maximum confidence \cite{croke,herzog1,herzog2} as limiting cases for sufficiently large values of $Q$. Moreover, it was found \cite{herzog3, bagan, nakahira} that when the maximum of $P_c$ is known as a function of the fixed value $Q$, then from this function one can also obtain the maximum of $P_c$ in  another optimal discrimination strategy where the error probability is fixed  \cite{touzel,hayashi-err,sugimoto}.  Optimal discrimination with a fixed rate $Q$  has been also investigated for a modified  optimization problem where in contrast to the usual assumption the prior probabilities  of the states are unknown \cite{nakahira1}.   

Solutions for optimal state discrimination with a fixed rate $Q$ of inconclusive results and with given prior probabilities of the states have been recently derived in three independent papers: Starting from the operator equations for the  optimality conditions, in Ref. \cite{herzog3} we obtained  analytical solutions for the discrimination of symmetric states and of two states occurring with arbitrary prior probabilities, where the two states are either pure or belong to a certain class of mixed qubit states. In  Refs. \cite{bagan} and \cite{nakahira} the respective authors showed that the optimization  problem with fixed $Q$ can be  solved by reducing it to  a  resulting minimum-error  problem the solution of which is known and to an additional optimization. They derived analytical solutions for discriminating between two pure states occurring with arbitrary prior probabilities and between the trine states \cite{bagan}, and also for the discrimination of geometrically uniform  states \cite{nakahira}.    

The present  paper goes beyond these previous investigations in several respects.  Based on the ideas of our earlier work \cite{herzog3} we study general properties of the optimal  measurement for discriminating with  fixed $Q$ between $N$  arbitrary qudit states in a $d$-dimensional Hilbert space. 
Specializing on the  case $d=2$, we develop a method for treating the optimal discrimination with fixed $Q$  between $N$ arbitrary qubit states, occurring with arbitrary prior probabilities. 
 In contrast to the previous papers \cite{herzog3,bagan, nakahira} we take into account  that often  the optimal measurement is not unique, which means that the maximal probability of correct results with fixed $Q$ can be obtained by a number of different measurements. 
In the special case $Q=0$, where the problem corresponds to minimum-error discrimination, our method  
differs from the approaches developed previously for treating the minimum-error discrimination of $N$ arbitrary qubit states \cite{samsonov, deconinck, kimura,  bae1}.
We obtain explicit analytical results for several problems that have not been solved before, mostly for the discrimination with fixed $Q$ between qubit states which posses a certain partial symmetry, but also for discriminating $N$ equiprobable qubit states and for the discrimination between a pure state and a uniformly mixed state in a $d$-dimensional Hilbert space.   
 
The paper is organized as follows. In Sec. II we start by considering the optimal discrimination with fixed $Q$ for $N$ qudit states. After this we specialize on the discrimination of $N$ qubit states in a two-dimensional Hilbert space and develop a method for solving the problem.
Sec. III is devoted to the  discrimination of $N$ partially  symmetric qubit states.    We conclude the paper in Sec. IV with discussing the relation of our method to  previous studies of  the minimum-error discrimination of  arbitrary qubit states and with a brief summary of results. The detailed derivations referring to Sec. III  are presented in the Appendix. 

\section{General theory}
\subsection{Optimal discrimination of qudit states}

We consider the discrimination between $N$ qudit states,  given by the
density operators $\rho_j$ ($j=1,\ldots, N$) and  prepared with the respective  prior
probabilities $\eta_j$, where $\sum_{j=1}^N \eta_j =1$.
 A complete  measurement performing the discrimination is  described by $N+1$ positive detection
 operators $\Pi_0, \Pi_1,\ldots, \Pi_N$ with  
\begin{equation}
\label {compl-0}
\Pi_0+\sum_{j=1}^N \Pi_j= I, \quad \Pi_0,\,\Pi_j \geq 0 \;\;{\rm for}\;\;j=1,\ldots, N,
\end{equation}
where $I$ is the identity operator in the  $d$-dimensional Hilbert space ${\cal H}_d$ jointly spanned by the
$N$ density operators $\rho_j$.
 The conditional probability that a quantum system prepared
 in the state $\rho_k$
 is inferred to be in the state $\rho_j$ is given by ${\rm Tr}(\rho_k\Pi_j)$, while ${\rm Tr}(\rho_k\Pi_0)$
is the conditional probability that the measurement yields an inconclusive outcome which does not allow to infer the state.
The overall probability of inconclusive results is then given by 
\begin{equation}
\label{Q}
Q=\sum_{j=1}^N\eta_j{\rm Tr\,} (\rho_j\Pi_0)= {\rm Tr\,}(\rho \Pi_0) \quad{\rm with}\;\;  \rho=  \sum_{j=1}^N \eta_j \rho_j,
\end{equation}
and the probability of correct results, $P_c$, reads
\begin{equation}
\label{Pc}
P_c  =\! \sum_{j=1}^N \!\eta_j{\rm Tr\,}(\rho_j\Pi_j)=1-Q-P_e,
\end{equation}
where $P_e$ denotes  the total rate of errors.
Our task is to  maximize $P_c$, or minimize $P_e$, respectively,  under the constraint that $Q$ is fixed at a given value.
Upon  introducing a Hermitian operator $Z$ and a scalar real amplifier $a$ the  necessary and sufficient optimality conditions take the form \cite{fiurasek,eldar1,herzog3}
\begin{eqnarray}
\label {optZ1}
(Z- a \,\rho)\;\Pi_0 =0, \;&& \! Z \!- a \rho  \geq  0,\\
\label {optZ2}
 ( Z  -\eta_j \rho_j )\Pi_j =0,&&    \! Z- \eta_j\rho_j  \geq  0\;\; \qquad
\end{eqnarray}
($j=1,\ldots,N$).
For $\Pi_0=0$ and $a=0,$  these conditions refer to minimum-error discrimination  \cite{helstrom,holevo}, where $Q=0$. 
 Provided that  Eqs. (\ref{optZ1}) and (\ref{optZ2}) are fulfilled, the rate of correct results takes its maximum value, given by
\begin{equation}
\label {Pcmax-}
 P_c^{max}\big|_Q
=\! \sum_{j=1}^N \!\eta_j{\rm Tr\,}(\rho_j\Pi_j)
={\rm Tr\,} Z-a Q.
\end{equation}
The detection operators $\Pi_j$ ($j=1,\ldots,N$) satisfying the optimality conditions need not always be unique, which means that different measurements can be optimal, yielding the same value  $P_c^{max}|_Q$.
This will be outlined later in this section and will be illustrated by examples in Sec. III.

It is useful to introduce  the ratio $R$ of the maximal rate of correct results, $P_c^{max}|_Q,$ and the rate of conclusive results, $1-Q.$
By taking the derivative we find with the help of  Eq. (\ref{Pcmax-}) that 
\begin{equation}
\label{lim3}
\begin{array}{ll} \frac{{\rm d} }{{\rm d}Q}R(Q) >0  \;& \mbox{if $ \; Z\neq a\rho,  $}\\
 \frac{{\rm d} }{{\rm d}Q}R(Q) =0  \;& \mbox{if $ \; Z= a\rho, $}
  \end{array}
\quad  {\rm where}\;\; R =  \frac{P_c^{max}|_Q}{1-Q}.  
\end{equation}
Here we used the fact that due to  the positivity constraint  $Z- a\rho\geq 0$ the condition $Z\neq a\rho$ is equivalent to ${\rm Tr} Z > a.$  
 Since ${\Pi}_0=I$ for $Q=1$, the operator  $\Pi_0$ must turn into a  full-rank operator when $Q$ approaches unity.  Eq. (\ref{optZ1}) therefore implies that  for sufficiently large values of $Q$ the condition $Z= a\rho$ is necessarily fulfilled \cite{fiurasek}. 
When for the given states the optimality conditions can be satisfied with   
\begin{equation}
\label{lim3a}
 Z\neq a\rho\quad {\rm if}\;\; Q \leq Q_u, \;\quad Z= a\rho \quad {\rm if}\;\; Q_u\leq Q\leq 1,
\end{equation}
then due to Eq. (\ref{lim3}) the ratio  $R$ grows with increasing $Q$ if $Q<Q_u$ and stays constant for $Q>Q_u$. Hence Eq. (\ref{lim3a}) defines $Q_u$   as the smallest value of $Q$ at which further increasing $Q$  does not yield any advantage, since the relative rate of correct results, given by $R$, remains constant. If $Z\neq a\rho,$  the operator $\Pi_0$ cannot be a full-rank operator, due to the equality condition in Eq. (\ref{optZ1}). For  $Z= a\rho,$ however, this condition does not put any restriction on the rank of $\Pi_0$. Eq. (\ref{lim3a}) takes into account that for $Q=Q_u$ the  expressions for $Z$ and $a$ determining the optimal solution are not unique, see the example at the end of Sec. II A.

\subsubsection{The limiting case of large $Q$}

Let us first review the  limiting case of large $Q,$ where $R$ does not change when $Q$ is increased, which means that $Z=a\rho.$  
 After substituting this expression for $Z$ into  Eq. (\ref{optZ2}) we obtain  the conditions  $(aI-\tilde{\rho}_j )\bar{\Pi}_j =0$ and $aI-\tilde{\rho}_j \geq 0$ for $j=1,\ldots,N$, where $\bar{\Pi}_j = \rho^{1/2}\Pi_j  \rho^{1/2}$ and $\tilde{\rho}_j=   \rho^{-1/2} \eta_j\rho_j   \rho^{-1/2}$. 
These conditions require that \cite{fiurasek,herzog3} 
\begin{equation}
\label{a-opt}
a= \underset{\overset{j}{}}{\mathrm{max}} \{C_j\}
 \quad{\rm with}\;\; C_j \!=\!{\rm max}\{{\rm eig }(\tilde{\rho}_j) \}.
 \end{equation}
The constant $C_j$, that is  the largest eigenvalue of $\tilde{\rho}_j$, characterizes  the maximum confidence  that can be achieved  for the individual measurement outcome $j$. In other words,  $C_j$ is equal to the maximum achievable value of the ratio  $\eta_j{\rm Tr\,}(\rho_j\Pi_j)/  {\rm Tr\,} (\rho\Pi_j)  $ which denotes the conditional probability that  the outcome $j$ is correct given it was obtained \cite{croke,herzog1,herzog2}. 
Due to Eq. (\ref{a-opt}) the conditions for optimality following from Eq. (\ref{optZ2}) require that $ \bar\Pi_j =0$ 
for states where $C_j <  a,$ that is where  the support of $ aI-\tilde{\rho}_j$ spans the whole Hilbert space
\cite{fiurasek,herzog3}.
This means  that  only the states for which  the maximum confidence $C_j$ is largest,   $C_j   = {\mathrm{max}}_j \{C_j\}$,    are guessed to occur in an optimal measurement where $Z= a\rho$, that is where $Q \geq  Q_u$ according to Eq.  (\ref{lim3a}). 
Using  Eq. (\ref{Pcmax-}) with $Z= a\rho$ and  $a   = {{\rm max}}_j \{C_j\}$ we arrive at  
\begin{equation}
\label{lim2}
P_c^{max}\big|_Q
= \underset{\overset{j}{}}{\mathrm{max}} \{C_j\}(1-Q)
\quad {\rm for} \;\;Q_u\leq Q \leq  1,
\end{equation}
where  $Q_u$ is defined by Eq.  (\ref{lim3a}). The actual value of $Q$ depends on the given states. In special cases where  $Q_u=0$ the maximum relative probability of correct results $P_c^{max}|_{Q}/(1-Q)$ cannot be increased at all by admitting inconclusive results but stays always equal to  $P_c^{max}|_{Q=0}$, see for instance the example of equiprobable states resolving the identity operator discussed in Ref. \cite {herzog3}.
   
If $Z=a\rho,$  the explicit expression for $Z$  can be easily determined with the help of  Eq. (\ref{a-opt}).  The equality conditions in  Eq. (\ref{optZ2}) then restrict the supports of the non-zero operators $\Pi_j$  to certain known subspaces.   $Q_u$ is the smallest value of $Q= 1-  \sum_{j}{\rm Tr}(\rho\Pi_j)$ that satisfies    
$\Pi_0= I-  \sum_{j}{\rm Tr}\Pi_j\geq 0$ on the condition that the operators $\Pi_j$ have their supports in the given subspaces.   $\Pi_0$ is a full-rank operator for $Q=1.$ When $Q$ decreases starting from 1, at a certain value  $Q_r$ one of the $d$ positive eigenvalues of $\Pi_0$ may become zero. If  an  eigenvalue immediately turns negative when $Q$ is decreased beyond  $Q_r$, then $Q_r$ is equal to $Q_u$. In this case $\Pi_0$ is  a full-rank operator in the whole region $Q_u<Q\leq1 $, see the  example at the end of Sec. II A, but this need not be valid in general.         

\subsubsection{The general case}

Now we exploit the optimality conditions without the restriction to the limiting case of large $Q$. If  $\Pi_0 \neq 0$, the  equality in Eq. (\ref{optZ1}) can be only fulfilled when the support of the operator $Z-a\rho$ does not span the full Hilbert space. This implies that at least one of its eigenvalues is equal to zero and the determinant therefore vanishes, yielding the condition
\begin{equation}
\label {det1}
 {\rm det} (Z  -a\rho)=0 \quad{\rm if} \;\;\Pi_0\neq 0.
 \end{equation}
Without lack of generality we suppose that in the optimal measurement the detection operators $\Pi_j$ are different from zero for $M$ states ($M\leq N$)  and vanish for the remaining $N-M$ states, which means that the remaining  states, if any, are never guessed to occur,  
\begin{equation}
\label {op}
\Pi_j \neq 0 \;\;{\rm if}\;\;j\!=\!1,\!\ldots,\!M, \qquad\Pi_j = 0 \;\;{\rm if}\;\;j\!=\!M+1,\!\ldots,\!N.
 \end{equation}
 In analogy to Eq. (\ref{det1}) we then arrive at the condition
\begin{equation}
\label {det2}
{\rm det} (Z  -\eta_j\rho_j)=0  \quad {\rm if}\;\;j=1,\ldots,M.
 \end{equation}
The requirement that the determinants in Eqs.  (\ref{det1}) and (\ref{det2}) vanish yields a system of $1+M$ real equations with  only $1+d^2$ real variables, namely  
the parameter $a$ and the $d^2$ real quantities determining the matrix elements of the Hermitean operator $Z$ acting  in  ${\cal H}_d$.   When the  quantum  states and their prior probabilities are completely arbitrary,   the system of equations therefore does not have a solution for $M > d^2$.   

In special cases, however, a  solution  
can  exist  where  more than $d^2$ states are guessed to occur,  for instance when the states have a certain symmetry. In these cases we  obtain the same  operator $Z$ and the same  parameter $a$ when  for some of the   states we drop Eq. (\ref{det2}),  keeping it for  $d^2$ or less states, and when we put the detection operators corresponding to the dropped states equal to zero, see the examples treated in Sec. III. 
The  measurement  then differs from the  one described by Eq. (\ref{op}), but yields the same value $P_c^{max}|_Q
={\rm Tr\,} Z-a Q.$  When an optimal measurement exists where more than  $d^2$ states are guessed, this  measurement is therefore not unique. 

The above considerations lead to the conclusion that  
we never need to make a guess for more than $d^2$ states in order to  discriminate optimally  between $N$ states in a $d$-dimensional
Hilbert space  with a fixed rate $Q$ of inconclusive results. 
This conclusion generalizes previous results  \cite{hunter,samsonov,deconinck} that have been obtained in a different way in the context of minimum-error discrimination, where $Q=0$. 
It follows that   for  discriminating optimally with fixed $Q$  between more than  $d^2$ arbitrary states we have to  consider  all possible subsets $S_K$  containing $d^2$ states separately, where $K=1,\ldots, {N \choose d^2}$, and we have to find the maximal rates of correct results  
$P_{c,K} ^{max}|_Q$ for discriminating the states within each  subset.  The largest of these  rates  then determines the optimal solution for discriminating the $N$ states,   in analogy to the findings obtained for the minimum-error discrimination of qubit states \cite {deconinck}.

The operator  $Z$ and the parameter $a$ satisfying  Eqs. (\ref{det1}) -  (\ref{det2}) only determine the optimal solution if  the optimality  conditions, Eqs. (\ref{optZ1}) and  (\ref{optZ2}), can be fulfilled  under the constraint that  $ \sum_{i=0}^M \Pi_i= I$. If for $M=d^2$ this is not the case,  we have to try whether for  $M=d^2-1$ a solution exists,  and so on  until $M=1$ where always the same state is guessed to occur when a conclusive outcome is obtained. The described procedure will be applied in Sec. II B for the case of qubit states.  

An interesting special  case arises when one of the states, say the first, has the  property that $\eta_1\rho_1-\eta_j\rho_j\geq 0$ for $j=2,\dots,N$, implying that $\eta_1 \geq \eta_j$. It is obvious that Eqs. (\ref{optZ1}) and  (\ref{optZ2})  and  also the completeness relation, Eq. (\ref{compl-0}),  are then fulfilled if    
\begin{equation}
\label {Ma}
Z= \eta_1\rho_1,\quad 
 \Pi_j=0 \;\; (j\!=\!2,\dots,N), \quad \Pi_1=I-\Pi_0 \geq 0.
\end{equation}
The explicit expressions for   $\Pi_0$  and $a$ follow from replacing $Z$ by 
$\eta_1\rho_1$ in Eq. (\ref{optZ1}),
 taking into account that ${\rm Tr}(\rho\Pi_0)=Q$ and that the positivity constraint $\eta_1\rho_1-a\rho \geq 0$ has to be satisfied.
The condition $I-\Pi_0 \geq 0$ restricts the validity of this solution to a certain region,
and for this region it follows from   Eq. (\ref{Pcmax-})  that  $P_c^{max}|_Q=\eta_1-aQ$, see the example treated below.   
Clearly, for minimum-error discrimination, where $\Pi_0=0$,  Eq. (\ref{Ma}) yields $\Pi_1=I$, which means that the maximum probability of correct results can be obtained without performing any measurement, simply by always guessing the state $\rho_1$ with the largest prior probability to occur \cite{hunter1,herzog0,herzog-bergou1}. 

In general,  it is very hard to obtain analytical solutions of the optimization problem with fixed $Q$ for $d>2.$  
In our previous paper \cite{herzog3} we derived the solution for a special class of linearly independent symmetric pure qudit  states.  In the following we treat another example, which contains also the special case discussed above.

\subsubsection{Discrimination of a uniformly mixed and a pure qudit state}

Let us consider the optimal discrimination between a uniformly mixed qudit state $\rho_1=\frac{1}{d}I$ and a pure qudit state  $\rho_2=|\psi\rangle\langle \psi|$, both living in the same Hilbert space  ${\cal H}_d$ and occurring with the prior probabilities $\eta_1$ and $\eta_2=1-\eta_1$, respectively. Taking into account that the spectral representation of  $\rho$ is given by  $\rho= \left(\frac{\eta_1}{d}+\eta_2\right) |\psi\rangle\langle \psi|  + \frac{\eta_1}{d}\left(I- |\psi\rangle\langle \psi|\right),$ where $I- |\psi\rangle\langle \psi|$ is a  $(d-1)$-dimensional  projector, 
 we find that the optimality conditions,  Eqs. (\ref{optZ1}) - (\ref{optZ2}), are satisfied for  $\frac{\eta_1}{d}\leq\eta_2$, or, equivalently,  for $\eta_2\geq \frac{1}{d+1}$,   if   
\begin{equation}
\label{umix2}
\begin{array}{ll} 
\Pi_1=I- |\psi\rangle\langle \psi|, \quad Z= \eta_2 |\psi\rangle\langle \psi|  + \frac{\eta_1}{d}\left(I- |\psi\rangle\langle \psi|\right), \\
\Pi_2= \left(1-  \frac{d}{\eta_1+d\eta_2}Q \right)   |\psi\rangle\langle \psi|, \quad a= \frac{d\eta_2}{\eta_1+d\eta_2},
 \end{array}
\end{equation}
 while for  $\eta_2\leq \frac{1}{d+1}$, corresponding to  $\eta_1\rho_1-\eta_2\rho_2 \geq 0$,  we obtain in agreement with  Eq. (\ref{Ma})   the solution
\begin{equation}
\label{umix3}
\begin{array}{ll} 
  \Pi_1\!=\!I\!-\! \frac{Qd}{\eta_1+d\eta_2}|\psi\rangle\langle \psi|,\;\; \Pi_2\!=\!0,  \;\; Z\!=\! \frac{\eta_1}{d}I,\;\;a= \frac{\eta_1}{\eta_1+d \eta_2}.
 \end{array}
\end{equation}
 In both cases we get  $\Pi_0=  \frac{d}{\eta_1+d\eta_2}Q |\psi\rangle\langle \psi|$. Since the detection operators have to be positive, this solution, where  $Z\neq a\rho $, only holds
 true  if $Q\leq \frac{\eta_1}{d}+\eta_2 = Q_u.$  
With the help of Eq. (\ref{Pcmax-}) we  find  that  
\begin{equation}
\label{umix1}
P_c^{max}\big|_Q \!=\!\left \{
\begin{array}{ll}\!\!  1-\frac{\eta_1}{d} - \frac {d \eta_2}{\eta_1+d \eta_2} Q  \; & \mbox{if $ \; \; Q\leq  Q_u,\;\;\eta_2\geq \frac{1}{d+1}   $}  \;\\
\!\!  \eta_1 - \frac{\eta_1}{\eta_1+d \eta_2}Q  \; & \mbox{if $
          \; \;Q\leq  Q_u,\;\;\eta_2\leq \frac{1}{d+1}   $},
  \end{array}
\right.
\end{equation}
For  $Q=0$ this result agrees with the solution  derived earlier for minimum-error discrimination \cite{herzog0,herzog-bergou1}. Note that  for  $\eta_2= \frac{1}{d+1},$ where $P_c^{max}|_Q= \eta_1-\frac{1}{2}Q,$ the optimal measurement  is not unique, as becomes obvious from comparing Eqs. (\ref{umix2}) and  (\ref{umix3}). 

It is easy to check that  the relative rate of correct results,  $P_c^{max}|_Q/({1-Q})$,  grows for both lines of  Eq. (\ref{umix1}) when the fixed rate $Q$ is increased, until it reaches unity for $Q=Q_u$. 
Since according to Eq. (\ref{a-opt}) the maximum confidences for discriminating $\rho_1$ and  $\rho_2$ are given by $C_1=1$ and $C_2=\frac{\eta_2 d}{\eta_1+\eta_2 d},$ respectively, Eq. (\ref{umix1}) yields    $P_c^{max}|_{Q_u}={{\rm max}}_j \{C_j\}(1-Q_u).$ Hence for $Q=Q_u$ the optimal measurement can be alternatively obtained when $Z=a\rho$ with $a={{\rm max}}_j \{C_j\}=1$, as becomes obvious from  Eq. (\ref{lim2}). 
Using Eqs. (\ref{optZ1}) - (\ref{optZ2}) we find that the optimality conditions are satisfied if  
\begin{equation}
\label{umix5}
\begin{array}{ll} 
\Pi_1= \frac{d(1-Q)}{(d-1)\eta_1}(I- |\psi\rangle\langle \psi|), \;\; \Pi_2=0,\quad Z=\rho,\;\; a=1. 
 \end{array}
\end{equation}
 Since the detection operators have to be positive, this result only holds true if $\Pi_0=I-\Pi_1\geq 0,$ that is  $Q\geq \frac{\eta_1}{d}+\eta_2 = Q_u.$  We therefore  arrive at   
\begin{equation}
\label{umix4}
P_c^{max}|_Q=1-Q\quad{\rm if}\;\;  Q_u\leq Q\leq 1,
\end{equation}
which because of Eq. (\ref{Pc}) expresses the fact  that errors do not occur in the optimal measurement for $Q\geq Q_u.$ Hence  in our example $Q_u$ is the smallest rate of inconclusive results for which an unambiguous discrimination is possible.   
 Eqs. (\ref{umix2}) or  (\ref{umix3}), as well as Eq. (\ref{umix5}), show that for $Q=Q_u$ the optimal measurement is a projection measurement with 
$\Pi_0=|\psi\rangle\langle \psi|$, $\Pi_1=I-\Pi_0$, and $\Pi_2=0$, which discriminates $\rho_1$ unambiguously, that is with the maximum confidence $C_1=1$, and  yields an inconclusive result when the state $\rho_2$ is present.   

While  the operator $\Pi_0$ is a  rank-one operator for $Q\leq Q_u$,  it follows  from   Eq. (\ref{umix5}) that $\Pi_0$ has the rank $d$  if $Q_u<Q\leq 1,$ that is in this example $\Pi_0$ turns into a full-rank operator at $Q=Q_u$.  As mentioned after Eq. (\ref{lim3a}), for $Q> Q_u$  the solution does not have any practical relevance since   the relative rate of correct results remains constant with growing $Q$ and its absolute rate $P_c^{max}|Q$ decreases. In the rest of the paper we do not provide the explicit expressions of the detection operators for $Q>Q_u.$

\subsection{Optimal discrimination of qubit states}

From now on we specialize our investigations on the optimal discrimination of qubit states in a two-dimensional joint Hilbert space  ${\cal H}_2$. 
Here it is sufficient to treat  the case where  the optimality conditions are satisfied with  $Z  -\eta_j \rho_j \neq 0$ ($j=1,\ldots,N$) and   $Z  -a \rho \neq 0.  $  
 When the first condition does not hold,  the optimal measurement is determined  by  Eq. (\ref{Ma}), while the violation of the second condition corresponds to the limiting case of large $Q$ where   Eq. (\ref{lim2}) applies. 

Following our earlier paper \cite{herzog3} we conclude that for  $Z \neq a \rho$ and $Z  \neq \eta_j \rho_j $ the equalities in the optimality conditions, Eqs. (\ref{optZ1}) and (\ref{optZ2}), imply that the non-zero optimal detection operators $\Pi_0$ and $\Pi_j$  are  proportional to the projector onto the eigenstates  $|{\pi}_0\rangle$ and $|{\pi}_j\rangle$ belonging to the eigenvalue zero of the operators   $Z  -a \rho $ and $Z  -\eta_j \rho_j $, respectively.
Hence if $\Pi_0 \neq 0$ and $\Pi_j \neq 0$ for $j\!=\!1,\!\ldots,\!M$ with $M\leq N$, see  Eq. (\ref{op}), the optimality conditions require that 
\begin{eqnarray}
\label {detec0}
\Pi_0=Q\frac{|\pi_0\rangle\langle \pi_0|}{\langle \pi_0|\rho |\pi_0\rangle} \,\,&& {\rm with}\quad
(Z-a\rho)\,|\pi_0\rangle=0, \\
\label {detec1}
 \Pi_j = \alpha_j |{\pi}_j\rangle\langle{\pi}_j|\;\;\; && {\rm with} \quad
(Z-\eta_j\rho_j)|{\pi}_j\rangle=0\qquad
\end{eqnarray}
$(j=1,\ldots,M)$, where $0<\alpha_j \leq 1$ and  $\langle\pi_0|\pi_0 \rangle\!=\! \langle \pi_j |\pi_j \rangle \!=\!1$.
 Since we supposed that  $\Pi_j=0$ for $j=M+1,\ldots,N,$  the completeness relation, Eq. (\ref{compl-0}), is given by     
\begin{eqnarray}
\label{compl}
&&\sum_{j=1}^M \alpha_j  |{\pi}_j\rangle\langle{\pi}_j|=I-Q\frac{|\pi_0\rangle\langle \pi_0|} {\langle \pi_0|\rho |\pi_0\rangle},\nonumber\\ &&{\rm with}\quad  \sum_{j=1}^M\! \alpha_j=2 -\frac{Q}{\langle \pi_0|\rho |\pi_0\rangle}.
  \end{eqnarray}
Due to Eqs. (\ref{detec0}) and  (\ref{detec1})  the projectors onto the normalized states $ |\pi_0^{\perp} \rangle$ and $|\pi_j^{\perp} \rangle$ that are orthogonal to $ |\pi_0 \rangle$ and $|\pi_j \rangle$,  respectively, are determined by the operator $Z$ and the parameter $a$ via the relations
\begin{equation}
\label {perp}
|\pi_0^{\perp} \rangle\langle \pi_0^{\perp}|=\frac{Z-a\rho}{{\rm Tr}Z-a}, \quad
|{\pi}_j^{\perp}\rangle\langle{\pi}_j^{\perp}|=\frac{Z\!-\!\eta_j\rho_j}{{\rm Tr}Z\!-\!\eta_j}.
\end{equation}
The  detection operators can be expressed  as    
\begin{equation}
\label {detec11}
\Pi_0= \frac{Q} {\langle \pi_0|\rho |\pi_0\rangle}({I-|\pi_0^{\perp} \rangle\langle \pi_0^{\perp}|}),\quad
\Pi_j = \alpha_j ( I-  |\pi_j^{\perp} \rangle\langle \pi_j^{\perp}|   )
\end{equation}
$(j=1,\ldots,M)$, where  ${\langle \pi_0|\rho |\pi_0\rangle}= 1-{\rm Tr}(\rho |\pi_0^{\perp} \rangle\langle \pi_0^{\perp}|)$. Using  Eq. (\ref{compl}) we find that the completeness relation takes the alternative form  
\begin{equation}
\label{compl1}
\sum_{j=1}^N \alpha_j
|{\pi}_j^{\perp}\rangle\langle{\pi}_j^{\perp}|
=I- 
 \frac{Q} {\langle \pi_0|\rho |\pi_0\rangle} |\pi_0^{\perp} \rangle\langle \pi_0^{\perp}|. 
  \end{equation}
For solving the optimization problem we have to  determine the parameter $a$ and the operator $Z$ which satisfy Eqs. (\ref{detec0}) -  (\ref{compl}), or, equivalently,  Eqs.  (\ref{perp}) - (\ref{compl1}), 
for a certain value of $M$ on the condition that the constants $\alpha_j$  are positive. For this purpose we can  proceed in two steps:

In the first step we use   Eq. (\ref{det1}) together with the positivity constraint  $Z\!  -a \rho \geq 0$  in order to  
 express $a$ and $\Pi_0$ in  dependence of the matrix elements of $Z,$ taken with respect to an orthonormal basis in ${\cal H}_2$.  It is advantageous to choose the particular  basis that  is the eigenbasis of $\rho$. Introducing the spectral representation 
\begin{equation}
\label {rho-diag}
 \rho = \sum_{j=1}^N\eta_j\rho_j= r|0\rangle\langle 0|+ (1-r)|1\rangle\langle1|,
\end{equation}
where $0<r<1$,  we represent the operator $Z$ as   
\begin{equation}
\label {Z}
Z = z_{00}  |0\rangle\langle 0|+ z_{11} |1\rangle\langle 1| +
z_{01} |0\rangle\langle 1|+z_{01}^{\star}|1\rangle\langle 0|.
\end{equation}
The requirement  ${\rm det} (Z\!  -a \rho )=0,$ see  Eq. (\ref{det1}),  then leads to
$(z_{00}-a r)\,[ z_{11}-a (1-r)]= |z_{01}|^2.$
 It turns out that from the two possible solutions only the result
\begin{equation}
\label{a-sol}
a=\frac{z_{00}}{2r}+\frac{z_{11}}{2(1-r)} -
\sqrt{\left[\frac{z_{00}}{2r}-\frac{z_{11}}{2(1-r)} \right]^2+\frac{|z_{01}|^2}{r(1-r)}}
\end{equation}
satisfies the inequality $z_{00}-a r \geq 0$ which implies that   also $Z\!  -a \rho \geq 0$   since the determinant vanishes.  After substituting this result for $a$ into  Eq. (\ref{perp}), we can express  the completeness relation of the detection operators, Eq. (\ref{compl1}), as a function of $Q$ and of  the
matrix elements of $Z$.     

In the second step we apply   Eqs. (\ref{det2}) and  (\ref{compl1}), together with the positivity constraints 
\begin{equation}
\label{constr}
Z-\eta_j\rho_j\geq 0 \;\;(j=1,\ldots,N),  \quad \alpha_j > 0\;\;(j=1,\ldots,M).
  \end{equation}
 The requirement ${\rm det} (Z  -\eta_j\rho_j)=0,$ valid for $j=1,\ldots,M$, leads to
\begin{equation}
\label{M-det}
(z_{00}-\eta_j\rho_{00}^{(j)})[z_{11}-\eta_j \rho_{11}^{(j)}]=| z_{10}-\eta_j \rho_{10}^{(j)}|^2 \quad(j=1,\ldots, M),
\end{equation}
where we used the representation  
\begin{equation}
\label {rhoj-ij}
\rho_j = \rho_{00}^{(j)}  |0\rangle\langle 0|+\rho_{11}^{(j)} |1\rangle\langle 1| +
\rho_{01}^{(j)} |0\rangle\langle 1|+\rho_{10}^{(j)}|1\rangle\langle 0|.
\end{equation}
After introducing the abbreviations
\begin{equation}
\label{arb1}
\Delta Z= z_{00} z_{11}- |z_{10}|^2,\quad \Delta \rho_j= \rho_{00}^{(j)}\rho_{11}^{(j)}- |\rho_{10}^{(j)}|^2
  \end{equation}
and after decomposing the non-diagonal matrix elements into their real and imaginary parts,  Eq. (\ref{M-det}) can be rewritten as
\begin{eqnarray}
\label{arb2}
&&\Delta Z + \eta_j^2  \Delta \rho_j  - \eta_j  \rho_{11}^{(j)}  z_{00} = F_j \qquad\qquad\qquad\\
\label{arb3} 
&&{\rm with}\;\;F_j\!=\!\eta_j \left( \rho_{00}^{(j)}  z_{11}-{\rm Re}\rho_{01}^{(j)}{\rm Re} z_{10}+{\rm Im}\rho_{01}^{(j)}{\rm Im} z_{10}\right),\nonumber
  \end{eqnarray}
which yields  for  $j=2,\ldots, M$  the equations
\begin{equation}
\label{arb4}
F_j - F_1=  \eta_j^2  \Delta \rho_j - \eta_1^2  \Delta \rho_1 -( \eta_j  \rho_{11}^{(j)} - \eta_1  \rho_{11}^{(1)}) z_{00}.
  \end{equation}
When the completeness relation, given by  Eq. (\ref{compl1}) together with  Eq. (\ref{perp}), is written in matrix form  we arrive at 
four real equations.    
Our task is to find  the  matrix elements of $Z$, determined by four real parameters, and the $M$ constants $\alpha_j$ which together fulfill the $M+4$ equations resulting from Eqs. (\ref{M-det}) and  (\ref{compl1}). Provided that the solution satisfies the positivity constraints     
given by Eq. (\ref{constr}), the optimization problem is solved. 

When the states are completely arbitrary, the system of $M$ equations given by  Eq. (\ref{M-det}) can only have a solution if $M\leq 4$. 
Therefore, when more than four completely arbitrary qubit states are to be optimally discriminated with fixed $Q$, we need to separately investigate  all possible subsets containing four states, as described  in  Sec. II for the discrimination of qudit states in ${\cal H}_d.$  We now focus on one such subset containing  $N=4$ qubit states  and discuss the different possible cases.  

{\it (i) Guessing of all four states is optimal}.
First we assume that $M=N=4$.  
 Eq. (\ref{arb4}) with  $j=2,3,4$  then represents 
a system of three linear equations for determining the variables on their left-hand sides,  $z_{11}$, ${\rm Re} z_{10}$ and
${\rm Im} z_{10}  $, as linear functions of the variable $ z_{00}$ occurring on the right-hand sides. If we insert the resulting expressions into Eq. (\ref{arb2}) with $j=1,$ we arrive at a quadratic equation for $z_{00}$. After solving this equation we get explicit expressions for the matrix elements of $Z$ and, due to Eq. (\ref{a-sol}), also for the parameter $a$. 
By inserting these  expressions into the four algebraic equations resulting from the matrix form of the completeness relation,  we arrive at a linear system of equations for explicitly determining the constants   $\alpha_1,\ldots  \alpha_4$.
Provided that the conditions in Eq. (\ref{constr}) are satisfied, we have succeeded in determining the optimal solution.  If one of the conditions is violated, we next investigate the case where $M=3$.  

{\it (ii) Guessing of three states is optimal.}
We now assume that $N=4$ and $M=3$, which means that in the optimal measurement one of the four states in the set, say the state with $j=4$, is never guessed to occur, corresponding to $\alpha_4=0$.  Using Eq. (\ref{arb4}) with $j=2,3$ together with Eq. (\ref{arb2}) we can determine
three of the four matrix elements of $Z$ in dependence of the remaining one, say $z_{00}$. After inserting the resulting expressions into the matrix form of Eq.  (\ref{compl1}) we again arrive at four equations. Using three of them, the constants $\alpha_1,\alpha_2$, and  $\alpha_3$ are obtained as functions of  $z_{00}$, and after inserting the results into the fourth we obtain an equation for determining $z_{00}$ which is, however, highly nonlinear for $Q\neq 0$. Provided that  the positivity conditions given by Eq. (\ref{constr}) with $j=1,2,3$ are satisfied, the solution has been obtained.

{\it (iii) Guessing of two states or of one state is optimal.} If the positivity conditions cannot be satisfied for  $M=4$ or $M=3$, a corresponding procedure has to be applied for  $M=2$. It may happen that also in this case the positivity conditions cannot be fulfilled and therefore $M=1$ in the optimal measurement, which means that it is  a projection measurement where either a particular state is guessed to occur or an inconclusive result is obtained.  
We mention at this place that in our previous paper \cite{herzog3} the explicit solution for discriminating two mixed qubit states with fixed $Q$ has been derived for cases where simple algebraic expressions result.

\subsubsection*{N equiprobable qubit states with equal purity}

The density operators of the qubit states can be represented in the general form
\begin{equation}
\label {rhoj-gen}
 \rho_{j} = p_j\,|\psi_j\rangle\langle \psi_j|+\frac{1-p_j}{2}\,I
 \quad(j=1,\ldots, N),
\end{equation}
where  $|\psi_j\rangle$ is a normalized  state vector, 
and where the parameters $p_j$ with $0 \leq p_j \leq 1$ depend on the purity of the respective states.
When the qubit states have equal prior probabilities and equal purities, that is when 
\begin{equation}
\label {equiprob}
\eta_j=\frac{1}{N}, \quad p_j=p \quad (j=1,\ldots, N),
\end{equation}
it is easy to check that  the operator   $ Z=\frac{1+p}{2N}I$ leads to  $Z-\frac{1}{N}\rho_j=\frac{p}{N}(I-|\psi_j\rangle \langle \psi_j|)$.
Hence the optimality condition given by Eq. (\ref{optZ2}) is satisfied if
\begin{equation}
\label {resolve2}
 \Pi_j=\alpha_j  |\psi_j\rangle\langle \psi_j| \quad {\rm with} \quad 0 \leq \alpha_j\leq 1
\end{equation}
for $j=1,\ldots,N$.  In order to exploit also the first optimality condition, Eq. (\ref{optZ1}), we use the spectral representation of   $\rho$, given by   Eq. (\ref{rho-diag}), where we denote its  eigenstates in such a way that   $r\geq  \frac{1}{2}$.
With $Z=\frac{1+p}{2N}\left(|0\rangle\langle 0|+  |1\rangle\langle 1|\right)$
it follows that
 Eq. (\ref{optZ1}) is fulfilled if
$a= \frac{1+p}{2N r}$ and $\Pi_0= \frac{Q}{r}|0\rangle\langle 0|$. Taking the completeness relation in the form of  Eq. (\ref{compl}) into account and  using Eq. (\ref{Pcmax-}), we find that the maximal probability of correct results is given by
\begin{eqnarray}
\label {Pmax}
&& \!\!P_c^{max}\big|_Q
=\frac{1+p}{N}\left(1-\frac{Q}{2r}\right),
\qquad\qquad\qquad\\
\label {compl-rel}
&& \!\!{\rm provided \; that } \;\; \sum_{j=1}^N \alpha_j  |\psi_j\rangle\langle \psi_j| + \frac{Q}{r}|0\rangle\langle 0 |=I\qquad
\end{eqnarray}
with $0\leq \alpha_j \leq 1$. Here $r$ denotes the largest eigenvalue of $\rho=\frac{1}{N}\sum_{j=1}^{N}\rho_j$, and $|0\rangle$ is the corresponding eigenstate. We note that for $Q=0$ and $p=1$  our result agrees with  the result obtained previously for the minimum-error discrimination of equiprobable pure states \cite{bae1}. 

When Eq. (\ref{compl-rel})  cannot be satisfied with non-negative values of $\alpha_j$ and  Eq. (\ref{Pmax}) therefore does not apply,   $\!P_c^{max}|_Q$  depends on the given states $|\psi_j\rangle$ not only via the eigenvalue $r$, but also in a direct way, as will become obvious  from the  example investigated in Sec. III C.  Moreover, even if for $Q=0$ a solution with non-negative coefficients $\alpha_j$ does exist, at least one of these coefficients may become negative if $Q$ exceeds a certain critical value $Q_{cr}$, which means that for $Q>Q_{cr}$ Eq.   (\ref{Pmax}) is not valid anymore, see Sec. III C.  
 We  mention  that in our previous paper \cite{herzog3} we considered  the  special case where $\frac{2}{N} 
\sum_{j=1}^N   |\psi_j\rangle\langle \psi_j|=I$, that is where $\rho=\frac{I}{2}$ and $r=\frac{1}{2}$, for which we obtained the complete solution     
$P_c^{max}\big|_Q
=\frac{1+p}{N}(1-Q)$, valid for arbitrary values of $Q$.

\section{Application to partially symmetric  qubit states}

\subsection{Properties of the partially symmetric  states} 

We wish to study a discrimination problem that yields simple analytical solutions but is nevertheless sufficiently general to demonstrate the basic features of  the method developed in Sec. II.   
For this purpose we consider the discrimination of  $N$ mixed qubit states, represented  by Eq. (\ref {rhoj-gen}), which 
fall into two groups containing $N_1$ and $N_2=N-N_1$ states, respectively. We suppose that within each group the qubit states have equal prior probabilities  and purities,
\begin{equation}
\label{rho-jj}
\eta_j, p_j  \!=\left \{
\begin{array}{ll}\!\!\eta,\,p    \,\; & \mbox{if $\;
              1\leq j \leq N_1  $}  \quad\\
\!\!\eta^{\prime}, p^{\prime} \; & \mbox{if $\;
             N_1 \!< \!j \!\leq\! N, $}\quad \\
 \end{array}
\right. N_1\eta+N_2\eta^{\prime}=1,
\end{equation}
and that $\rho_{j} = p_j\,|\psi_j\rangle\langle \psi_j|+\frac{1-p_j}{2}$ with 
\begin{equation}
\label{psi-j}
|\psi_j \rangle \!=\left \{
\begin{array}{ll}\!\! \sqrt{b}\, |0\rangle \!+\! e^{i\phi_j}\sqrt{1-b}\, |1\rangle  \quad \mbox{if $\;
              1\leq j \leq N_1  $}\\
\!\!\sqrt{c}\,|0\rangle \!+\! e^{i\phi_j}\sqrt{1-c}\,|1
\rangle  \quad \mbox{if $\;
               N_1 \!< \!j \!\leq\! N,  $}
 \end{array}
\right.\quad
\end{equation}
where $0 < b < c$.  The diagonal density matrix elements $\rho_{00}^{(j)}$ of  $\rho=\sum_{j=1}^N \eta_j\rho_j$ are then given by
\begin{eqnarray}
\label{rho-j}
\rho_{00}^{(j)}   =\left \{
\begin{array}{ll} p b+\frac{1}{2}(1-p) \equiv s   \,\; & \mbox{if $\;
              1\leq j \leq N_1  $}  \quad\\
 p^{\prime} c+\frac{1}{2}(1-p^{\prime}) \equiv s^{\prime}  \, \; & \mbox{if $\;
             N_1 \!< \!j \!\leq\! N. $}\quad \\
 \end{array}
\right.
\end{eqnarray}
In addition,  we assume  that $N_1\geq 2$, and we consider the cases where 
\begin{eqnarray}
\label{zero-i}
 \sum_{j=1}^{N_1} e^{i\phi_j}=0 \; \;{\rm and}\;\; (\rm { i})&&   \!\! \sum_{j=N_1+1}^{N_1+N_2}\! e^{i\phi_j}\!=0\;\;(N_2 \geq 2), \;\;{\rm or} \qquad\nonumber\\
\label{zero-ii}
  (\rm  {ii})&&  \!\! N_2 =1,\;\; c=1, \;\;{\rm or}\nonumber\\
\label{zero-iii}
(\rm { iii})&&  \!\! N_2 =0.
\end{eqnarray}
 Eqs.  (\ref{rho-jj}) -  (\ref{zero-i}) imply that  the non-diagonal density matrix elements of  $\rho=\sum_j\eta_j\rho_j$ vanish and yield the spectral representation  
\begin{equation}
\label {rhodiag}
 \rho=r  |0\rangle\langle 0|+(1-r) |1\rangle\langle 1|
\quad
{\rm with}\;\;
r= N_1 \eta  s +N_2 \eta^{\prime}s^{\prime}.
\end{equation}
\begin{figure}[t!]
\center{\includegraphics[width=7 cm]{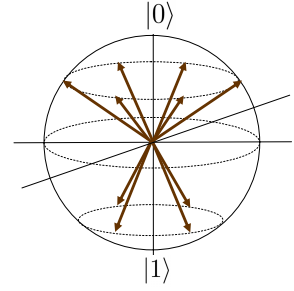} }
\caption{Bloch vectors for an example of partially symmetric pure states ($p=1$) with $N_1= 4$, $b < 0.5$ and $N_2 = 6$, $c  > 0.5.$ }
\end{figure}
The state vectors $|\psi_j\rangle$ in Eq. (\ref{psi-j}) can be visualized with the help of a Bloch sphere where  $|0\rangle$ and  $|1\rangle$ are the north and south pole, respectively, that is
$|\psi_j\rangle = \cos\frac{\theta_j}{2}|0\rangle + e^{i\phi_j}\sin\frac{\theta_j}{2}|1\rangle$
with $0\leq \theta_j \leq \pi$ and $0\leq \phi_j \leq 2\pi$. 
Clearly,  in each group the end points of the Bloch vectors representing the underlying states $|\psi_j\rangle$ have  the same latitude,  lying in the upper hemisphere if $b, c > \frac{1}{2}$ and in the lower hemisphere if $b, c < \frac{1}{2}$, see Fig. 1.
The phases can be represented as
\begin{equation}
\label{phi}
\phi_j  \!=\left \{
\begin{array}{ll}\!\! {2\pi i}\frac{j-1}{N_1} \;\,\quad\quad \quad \quad \mbox{if $\;
              1\leq j \leq N_1  $}\\
\!\! \delta+  {2\pi i}\frac{j-N_1-1}{N_2}  \quad \mbox{if $\;
               N_1 \!< \!j \!\leq\! N,  $}
 \end{array}
\right.\quad
\end{equation}
where $\delta$ is an arbitrary  phase shift.  Clearly,  the density operators obey the symmetry conditions
\begin{equation}
\label{rho-symm}
\rho_j  \!=\left \{
\begin{array}{ll}\!\! U^{(j-1)}\rho_1 U^{\dag( j-1) } \qquad\quad \quad \qquad \mbox{if $\;
              1\leq j \leq N_1  $}\\
\!\! V^{(j-N_1-1)}\rho_{N_1+1} V^{\dag(j-N_1-1) }  \quad \mbox{if $\;
               N_1 \!< \!j \!\leq\! N,  $}
 \end{array}
\right.\quad
\end{equation}
where  the unitary operator $U=|0\rangle\langle 0|+ {\rm exp}\frac{2\pi i}{N_1}|1\rangle\langle 1|$ with $U^{N_1}=I$
characterizes the symmetry  in the first group of states, and   $V=|0\rangle\langle 0|+ {\rm exp}\frac{2\pi i}{N_2}|1\rangle\langle 1|$ with $V^{N_2}=I$ refers to the second group.  In the case (ii) we put $V=I$. The cases (i) and ( ii) belong to the specific kind of partially symmetric states introduced in our earlier paper, Ref. \cite{herzog3},
 while in the case (iii) the $N$ states are symmetric.

For later purposes we still calculate the maximum confidence $C_j$ for discriminating an individual state $j$, using Eq. (\ref{a-opt}). We get  $C_j=C $ for $ 1\leq j \leq N_1$, and  $C_j=C^{\prime}$  for $N_1 <j\leq N$, where
\begin{equation}
\label {maxconf1}
C = \eta\frac{r-s(2r-1)}{2r(1-r)}\left[1+\sqrt{1-\frac{(1-p^2)r(1-r)}{[r-s(2r-1)]^2}}\right],
\end{equation}
and where  $C^{\prime}$ follows from  Eq. (\ref{maxconf1}) when
$s, p$, and $\eta$ are replaced by $s^{\prime}, p^{\prime}$, and $\eta^{\prime}$, respectively.  In  the special case where $N_2=0$, meaning that  $\eta=1/N$ and  $r=s,$ the expression for $C$ is in agreement with our earlier result \cite{herzog2} for the maximum-confidence discrimination of symmetric mixed qubit states.

\subsection{Derivation of the optimal measurement}

In order to solve our optimization problem we  apply the method outlined in  Sec. II. 
Due to the partial symmetry of the states, the treatment can be considerably  simplified. In fact, 
assuming that at least from one of the two groups, say the first, two or more states are guessed to occur, we get  for these states  from   Eq. (\ref{M-det}) the conditions   $(z_{00}-\eta s)[z_{11}-\eta(1-s)]=| z_{10}-\eta \rho_{10}^{(j)}|^2 $  with  $\rho_{10}^{(j)}= p\sqrt{b(1-b)}e^{i\phi_j}.$  Since these conditions can only be simultaneously satisfied for two or more different phases $\phi_j$ if $z_{10}=0$, we have to search for an operator $Z$ where  
\begin{equation}
\label {Z}
 Z = z_{00}  |0\rangle\langle 0|+ z_{11} |1\rangle\langle 1|, \quad \mbox{that is}\;\; [Z,\rho]=0.
\end{equation}
We note that the requirement  $[Z,\rho]=0$ was derived already in our earlier paper \cite{herzog3}, using the fact that there always exists an optimal measurement where the detection operators $\Pi_j$ obey the same symmetry as the  density operators $\rho_j$  of the partially symmetric states.

 As mentioned in the beginning of Sec. II B,  it is sufficient to perform the derivation for the case where $Z\neq a\rho$ and $Z\neq \eta_j\rho_j$ $(j=1,\ldots,N)$. In the first step we make use of Eq. (\ref{det1}). Since  $z_{10}=0$,  the condition ${\rm det}(Z-a\rho)=0$ is equivalent to
$(z_{00}-a r)\,[ z_{11}-a (1-r)]= 0$.
Taking the positivity constraint $Z-a\rho \geq 0$ and the relation ${\rm Tr}(\rho\Pi_0)=Q$
 into account, we find that the optimality condition in Eq. (\ref{optZ1}) is satisfied if 
\begin{equation}
 \label{Pcmax1}
\;a\!= \!{\rm min}\left\{\! \frac{z_{00}}{r},\frac{z_{11}}{1-r} \!\right\},\,\;\; \Pi_0\!=\!\left \{
\begin{array}{ll}\!\! \frac{Q}{r}\, |0\rangle\langle0| \quad \,\,\;\;\mbox{if $\;
            \frac{z_{00}}{r}< \frac{z_{11}}{1-r}  $}\\
\!\! \frac{Q}{1-r}\, |1\rangle\langle1| \quad
         \mbox{if $\;
            \frac{z_{00}}{r}> \frac{z_{11}}{1-r} $,}
 \end{array}
\right.\quad
\end{equation}
where the expression for $a$ is in accordance with  Eq. (\ref{a-sol}). Note that  if   $\frac{z_{00}}{z_{11}}= \frac{r}{1-r}$ both eigenvalues of  $Z-a\rho$ vanish which means that  $Z=a\rho$ and  which corresponds to the case $Q\geq Q_u$, yielding Eq.  (\ref{lim2}).   
Due to  Eq. (\ref{Pcmax-}) the maximum probability of correct results for  $Z\neq a\rho,$ that is for  $Q \leq Q_u,$ takes the form
\begin{equation}
\label {sol}
 P_c^{max}\big|_Q= \left \{
\begin{array}{ll}
\!\! {z}_{00}  ( 1 - \frac{Q}{r} )  +{z}_{11} \quad \;\,\;\;\mbox{if $\;
             \frac{z_{00}}{r}< \frac{z_{11}}{1-r} $}\\
\!\! {z}_{00}   +{z}_{11}  (1-\frac{Q}{1-r}) \, \quad
         \mbox{if $\;
            \frac{z_{00}}{r}> \frac{z_{11}}{1-r}. $}
 \end{array}
\right.\quad
\end{equation}
 
In the second step we apply  Eq. (\ref{det2}) and the completeness relation, represented by Eq. (\ref{compl1}), together with the positivity constraints given by Eq. (\ref{constr}).   
Using   Eq. (\ref{M-det})  we arrive at  the conditions
\begin{equation}
\label {two1a}
\begin{array}{ll}
(z_{00}-\eta s)\,[z_{11} -\eta(1-s)]=\eta^2 p^2\, b(1\!-b)\,\;\;\;{\rm if}\;A_1>0,\qquad\\
(z_{00}\!-\!\eta^{\prime} s^{\prime})[z_{11}\! -\!\eta^{\prime}(1\!-\!s^{\prime})]=\eta^{\prime 2}p^{\prime 2}c(1\!-\!c)\;\;\,\,{\rm if}\;A_2>0,\quad
\end{array}
\end{equation}
where we introduced 
\begin{equation}
\label{AA}
 A_1\!=\!\sum_{j=1}^{N_1} \alpha_j,\;\;  A_2\! =\!\!\!\! \sum_{j=N_1+1}^N\!\!\!\!\alpha_j\;\;\;{\rm with}\;\; \alpha_j \geq 0\;\;( j=1,\ldots,N).
\end{equation}
Because of   Eq. (\ref{detec1})  the condition  $A_1>0$ is equivalent to requiring that at least one state from the first group is guessed, and  $A_2>0$ means that at least one state from the second group is inferred to occur. 
 Next we make use of Eqs.  (\ref{compl1}) and  (\ref{perp}).  Because of the explicit form of the operator $\Pi_0$, given by Eq. (\ref{Pcmax1}),  the diagonal matrix elements of  Eq.  (\ref{compl1})  lead to the conditions  
\begin{eqnarray}
\label{compl2a}
A_1 \frac{z_{00}-\eta s}{{\rm Tr}Z-\eta} + A_2 \frac{z_{00}-\eta^{\prime} s^{\prime}}{{\rm Tr}Z-\eta^{\prime}}\!=\!f_0,\qquad\qquad\qquad \\
\label{compl2b}
A_1 \frac{z_{11}-\eta (1-s)}{{\rm Tr}Z-\eta} + A_2 \frac{z_{11}-\eta^{\prime}(1- s^{\prime})}{{\rm Tr}Z-\eta^{\prime}}\!=\!f_1,\quad\;
\end{eqnarray}
 where the constants $f_0$ and $f_1$ read 
\begin{equation}
\label {f}
\begin{array}{ll}
f_0=1,\;\;  f_1=1-\frac{Q}{r}\;\; \quad\,\;\;\mbox{if $\;
             \frac{z_{00}}{r} < \frac{z_{11}}{1-r}\, , $}\\
f_1=1,\;\;f_0=1\!-\!\frac{Q}{1-r}   \, \quad\;\;\,
       \mbox{if $\,\,
         \frac{z_{00}}{r} > \frac{z_{11}}{1-r}\,  $}.
\end{array}
\end{equation}
Since $z_{10}=0,$ the non-diagonal elements of   Eq.  (\ref{compl1}) yield the condition
\begin{equation}
\label{compl2c}
\eta p\frac{\sqrt{b(1\!-\!b)}}{{\rm Tr}Z-\eta}\sum_{j=1}^{N_1} \alpha_j \,e^{i\phi_j} +\eta^{\prime}p^{\prime}\frac{\sqrt{c(1\!-\!c)}}{{\rm Tr}Z-\eta^{\prime}}\!\!\!\sum_{j=N_1+1}^{N}\! \!\!\alpha_j \,e^{i\phi_j}\! =\!0. \qquad
\end{equation}
 Due to the symmetry properties of the partially symmetric states, Eq. (\ref{compl2c}) is certainly fulfilled if   
\begin{equation}
\label {barA}
\alpha_j\!=\!\frac{{A}_1}{N_1}\quad {\rm for} \;\, 1\leq j\leq N_1, \quad \alpha_j\!=\!\frac{{A}_2}{N_2} \quad{\rm for} \;\, N_1\!<\!j\!\leq \!N, 
\end{equation}
that is if in the optimal measurement the detection operators for the individual states obey the same symmetry as their  density operators, as can be seen with the help of Eqs. (\ref{detec11}) and  (\ref{perp}).  
However, other choices of the coefficients $\alpha_j$ and therefore of the detection operators are also possible.  Taking into account that $\phi_j=  {2\pi i}\frac{j-1}{N_1}$ for $1\!\leq\! j \!\leq \!N_1,$  see Eq. (\ref{phi}), we find for instance  that  $\sum_{j=1}^{N_1} \alpha_j \,e^{i\phi_j}=0$ if for even numbers $N_1$  
\begin{equation}
\label {even}
\alpha_1=\alpha_{\frac{N_1}{2}+1}=\frac{A_1}{2},
 \quad \alpha_j= 0\;\; {\rm else }\;\;(1\!\leq\! j \!\leq \!N_1)  
\end{equation}
and for odd numbers $N_1$
\begin{eqnarray}
\label {odd}
\alpha_1\!=\!\alpha_2\! = \!\frac{A_1}{2+2\cos\frac{\pi}{N_1}},\quad \alpha_{\frac{N_1+3}{2}}= \frac{A_1 \cos\frac{\pi}{N_1}}{1+\cos\frac{\pi}{N_1}},\quad
\end{eqnarray}
while  all other coefficients $\alpha_j$ $(1\!\leq\! j \!\leq \!N_1)$ vanish.
Similarly,   various  solutions with     $\sum_{j=N_1+1}^{N} \alpha_j \,e^{i\phi_j}=0$  exist  for the coefficients $\alpha_j$  that belong to the states in the second group and depend on $A_2$. Moreover, if $A_1, A_2 >0$  Eq. (\ref{compl2c}) may be also solved in a way where the terms referring to the two groups of states do not vanish separately. 

In order to find an optimal measurement  
we have to determine the values of  $z_{00}$, $z_{11}$, $A_1$ and $A_2$ that  satisfy Eqs. (\ref{two1a}) - (\ref{f}) with 
 $A_1\geq 0$ and $A_2\geq 0$, 
where  according to Eq. (\ref{optZ2}) the  positivity constraints   
\begin{equation}
\label{constr1}
Z-\eta\rho_j \geq 0\;\; {\rm if}\;1\!\leq\! j \!\leq \!N_1,\quad  Z-\eta^{\prime}\rho_j \geq 0\;\; {\rm if}\;N_1\! < \! j \!\leq \!N 
\end{equation}
have to be fulfilled.  For $A_1, A_2 >0$, where both lines of   Eq. (\ref{two1a}) apply, the positivity constraints are satisfied provided that   
\begin{equation}
\label {constr2}
z_{00}-\eta s\geq 0, \quad  z_{00}-\eta^{\prime} s^{\prime}\geq 0,
\end{equation}
while for $A_1>0$  and $A_2=0$ the additional condition 
\begin{equation}
\label {constr2a}
 (z_{00}-\eta^{\prime} s^{\prime})\,[z_{11} -\eta^{\prime}(1-s^{\prime})]\geq \eta^{\prime 2} p^{\prime 2} c(1-c)
\end{equation}
has to be taken into account. A similar condition arises when  $A_2>0$ and $A_1=0.$
Since  in general the resulting solutions  are rather  involved,
we treat the general case only in implicit terms and provide explicit derivations of special solutions in the Appendix. Two different cases have to be considered:

{\it (i) Guessing states from both groups is optimal.}
For   $A_1,A_2>0$  Eq. (\ref{two1a}) represents a system of two coupled equations from which we obtain      
 expressions for  $z_{00}$ and $z_{11}$  after  solving a quadratic equation. 
When these expressions are inserted   
 into Eqs.  (\ref{compl2a}) and  (\ref{compl2b}),  the resulting linear system of equations yields the values of $A_1$ and $A_2$,  which depend on $Q$ due to Eq. (\ref{f}).
Provided that these values are indeed  positive and that Eq. (\ref{constr2}) is satisfied, we have obtained the optimal solution. If this is not the case, a measurement where states from both groups are guessed cannot be optimal.   
  
{\it (ii) Guessing   states from only one group is optimal.}
To be specific, let us assume that only the states from the first group are guessed,  which means that  $A_1>0$ and $A_2=0$. 
From  Eqs. (\ref{compl2a}) and  (\ref{compl2b}) with  $A_2=0$  we  obtain the two equations    
\begin{equation}
\label{first}
A_1=f_0 + f_1, \quad  {f_1}(z_{00}-\eta s)={f_1}\left[z_{11}-\eta(1- s)\right].
\end{equation} 
Using in addition the first line of  Eq. (\ref{two1a}) and taking  Eq. (\ref{f}) into account, we arrive at three equations for determining   $z_{00},  z_{11}$ and $A_1$  in dependence of $Q$.
Provided that indeed $A_1>0$ and that  Eqs. (\ref{constr2}) and  (\ref{constr2a})  are satisfied,  the optimal solution has been determined.  
For the case where only the states from the second group are guessed we can proceed in an analogous way.

\subsection{ Equiprobable partially symmetric states with equal purity}

In the next two subsections we present the complete  solutions  for two simplified but non-trivial discrimination problems for partially symmetric states. In our first problem we suppose that the purity of the  states in the two groups is the same,  $p=p^{\prime}$, and that the states occur with the  same prior probability, $\eta=\eta^{\prime}=\frac{1}{N}$.  We denote the eigenstates of $\rho$ in such a way that  $r$, given by Eq. (\ref{rhodiag}), is the largest eigenvalue,
\begin{equation}
\label{r1}
r=  \frac{1}{2}+\frac{p}{2N} \left[N_1(2b-1)+N_2(2 c-1)\right] \geq \frac{1}{2}.
\end{equation}
To be specific, we suppose that  $c>b$  which implies that  $c\geq \frac{1}{2}$ if $r\geq \frac{1}{2}$.
With the help of  Eq. (\ref{maxconf1}) this yields the relation $ C \geq C^{\prime}$ for the maximum achievable confidences for discriminating the states in the two groups, where the equality sign holds if $r=\frac{1}{2}$.
As derived in the Appendix, the maximum probability  of correct results with fixed $Q$  is then given  by the  solution   
\begin{eqnarray}
\label{solQ}
P_c^{max}\big|_Q \!=\!\left \{
\begin{array}{ll}\!\!  P_{I}(Q) \; & \mbox{if $\;
          \; \;0\;\,\leq Q \leq Q_{cr}   $}\, \;\,  \quad\\
 \!\!P_{II}(Q) \; & \mbox{if $\;
               Q_{cr}\leq Q  \leq Q_u $}\;\;\, \quad\\
\!\! C(1-Q) \;\,\, & \mbox{if $\;\;
               Q_u\leq Q \leq 1 ,$} \quad\, 
 \end{array}
\right.
\end{eqnarray}
where $Q_{cr}= r \frac{1-2b}{1-b} $ and  $P_I(Q)=  \frac{1+p}{N}\left(1-\frac{Q}{2r}\right)$,  in agreement with   Eq.  (\ref{Pmax}),
while  $Q_u$ and  $P_{II}(Q)$   are  defined by Eqs.  (\ref{solQ3}) and (\ref{A-2}) in the Appendix, respectively.
Three different regions of $Q$ have to be distinguished: 
\begin{figure}[t!]
\center{\includegraphics[width=7.5 cm]{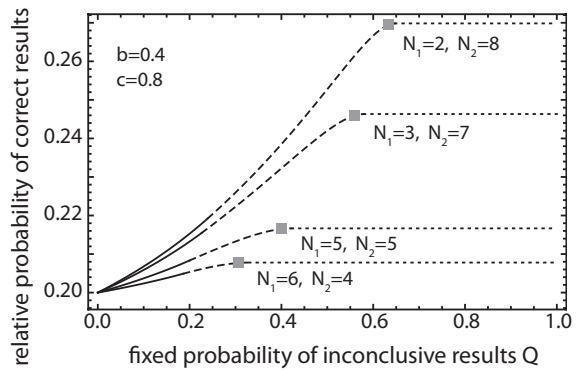} }
\caption{Relative probability of correct results $\frac{P_c^{max}|_Q}{1-Q}$ versus the fixed probability $Q$ of inconclusive results in an optimal measurement for discriminating {\it N} = 10  equiprobable partially symmetric pure states ({\it p} =1, {\it b} = 0.4, {\it c} = 0.8) for different values of $N_1$ and $N_2.$}
\end{figure}

For $ 0\leq Q< Q_{cr}$, corresponding to the full lines in Fig. 2,  states from both groups are guessed in the optimal measurement.  Clearly, this  only applies if  $Q_{cr} > 0$, that is if  $b< \frac{1}{2}$.
 $P_c^{max}|_Q$ is then determined by   $P_I(Q)$, in agreement with the general result  derived in  Eq. (\ref{Pmax}).
  In this region  the optimum  detection operators are $\Pi_0=\frac{Q}{r}|0\rangle\langle0|$ and
$ \Pi_j =  \alpha_j |{\psi}_j\rangle\langle{\psi}_j|$ for  $1\!\leq\! j\! \leq \!N$,
 where the states $|\psi_j\rangle$ are given by Eq. (\ref{psi-j}). The values of the constants $\alpha_j$ can be determined by applying the method  described in Sec. III B,  using the explicit expressions for $A_1$ and $A_2$ 
given in the Appendix.

In the region where $Q_{cr}\leq Q \leq Q_u$, corresponding to the dashed lines in Fig. 2, 
only the states of the first group are guessed in the optimal measurement. The optimal detection operators follow from  the expression for  $Z$, see Eq. (\ref{z0}), and from  Eqs. (\ref{perp}) and (\ref{detec11}). The constants $\alpha_j$  can be obtained as described Sec. III B,  with the help of the  value of $A_1$ given by Eq. (\ref{A1}). 
If $b>\frac{1}{2}$ and therefore  $Q_{cr} < 0$ the second line of  Eq. (\ref{solQ}) holds true in the whole region $0\leq Q\leq Q_u$. Since $c>b$, the Bloch vectors of the states are then confined to the upper hemisphere  of the Bloch sphere.   

The  region  where $Q\geq Q_u$ corresponds to the limiting case of large $Q$ described in Sec. II A. In  this region   the ratio 
$P_c^{max}|_Q/(1-Q)$, does not increase anymore with growing $Q$ but stays equal to $C$, see the dotted lines in Fig. 2. The squares in Fig. 2 indicate the points where $Q=Q_u$ for different numbers  of $N_1$ and $N_2$.

Two special cases are worth mentioning. For $Q=0$, that is for minimum-error discrimination, we obtain from Eq. (\ref{solQ}) 
\begin{equation}
\label{min-err}
P_c^{max}\big|_{Q=0} \!=\left \{
\begin{array}{ll}\!\!  \frac{1+p}{N}  \; & \mbox{if $\;
          c \geq \frac{1}{2}, \;\,b< \frac{1}{2}\;\,   $} \\
\!\! \frac{1+2p\sqrt{b(1-b)}}{N}\;  & \mbox{if $\;
               c >b\geq \frac{1}{2}. $}\; 
 \end{array}
\right.
\end{equation}
When the first line applies,  states from both groups are guessed in the optimal measurement, while otherwise only states from the first group are guessed to occur.   
Interestingly, the solution only depends on the total number $N=N_1+N_2$ of the states and is independent of their distribution over the two groups. 

The second special case refers to  $N_2=0$, that is  to the discrimination of $N_1=N$ equiprobable symmetric mixed qubit states.   Eq. (\ref{r1}) then reduces to   $r=s \geq \frac{1}{2}$ and requires that  $b\geq \frac{1}{2}$.  This means that the first line  of Eq. (\ref{solQ}) does not apply, and that $P_{II}(Q)$ is determined by  the first line of  Eq. (\ref{A-2}).  This solution   coincides with our earlier result \cite{herzog3} that was derived for symmetric detection operators. In contrast to this, the present derivation does not impose any restriction on the choice of the detection operators. In particular, it shows that for $N$ symmetric mixed qubit states  an optimal discrimination with fixed $Q$ can be always accomplished when not more than three  (for $N$ odd) or two (for $N$ even) states are guessed to occur, as follows from Eqs. (\ref{even}) and (\ref{odd}).  For  $Q=0$, that is for minimum-error discrimination,  the optimum measurement therefore can be always realized by a simple projection measurement if $N$ is even.

\subsection{Partially symmetric pure states with different  prior probabilities}

Now we assume that the states are pure,  $p=p^{\prime}=1$,  and that the respective prior probabilities $\eta$ and $\eta^{\prime}$ in the two groups of states can be different, where  $N_1\eta+N_2\eta^{\prime}=1$.  
In order to obtain the complete solution for arbitrary values of $\eta$ we   have to consider both the cases  where the relations $C\geq C^{\prime}$ and  $C \leq  C^{\prime}$ hold true between the maximum confidences for the states in the two groups, see Eqs. (\ref{pure4a}) and  (\ref{Pcmax2}) in the Appendix.  

We present explicit results for  the simplest  case, where  $N_1 =2$,   $N_2=1$ and  $c=1$, that is where the states are given by   
\begin{equation}
\label{mirror}
|\psi_{1/2}\rangle=\sqrt{b}|0\rangle \pm \sqrt{1-b}|1\rangle, \quad |\psi_3\rangle=|0\rangle
\end{equation}
and occur with the prior probabilities  $\eta$ for  $|\psi_{1}\rangle$ and $|\psi_{2}\rangle$, and $\eta^{\prime}=1-2\eta$ for $ |\psi_{3}\rangle,$ respectively.  
These states correspond to the three mirror-symmetric pure states the minimum-error discrimination of  which has been previously investigated \cite{andersson}. 
The condition  $C\geq C^{\prime}$ holds true provided that  $\eta \geq \frac{1}{2+4b}\equiv \eta_0$, 
where for $\eta=\eta_0$ we get $C=C^{\prime}=2/3.$ 
As shown in the Appendix,  for $C\geq C^{\prime}$ the maximum probability of correct results reads  
\begin{eqnarray}
\label{solQ1}
P_c^{max}\big|_Q \!=\!\left \{
\begin{array}{ll}\!\!  P_{I}(Q) \; & \mbox{if $\;
          \; \;0\;\,\leq Q \leq Q_{cr}   $}\, \;\,  \quad\\
 \!\!P_{II}(Q) \; & \mbox{if $\;
               Q_{cr}\leq Q  \leq Q_u $}\;\;\, \quad\\
\!\! C(1-Q) \;\,\, & \mbox{if $\;\;
               Q_u\leq Q \leq 1 ,$} \quad\, 
 \end{array}
\right.
\end{eqnarray}
where $Q_u$ is  determined by Eqs. (\ref{solQ3}) and (\ref{solQ2a}) with $p=1$, and where  $Q_{cr}$,  $P_I(Q)$, and  $P_{II}(Q)$ are given  by  Eqs. (\ref{crit}) - (\ref{mirror1a}) with  $r= 1-2\eta (1-b)$.
On the other hand, for $C\leq C^{\prime}$ we obtain   
\begin{eqnarray}
\label{pure3prime}
P_c^{max}\big|_Q =\left \{
\begin{array}{ll}
  P_{II}(Q) \; & \mbox{if $\;
       \;  \; \; \;0\;\,\leq Q\leq Q_{cr}^\prime   $} \;\,  \quad\\
  P_{I}^{\prime}(Q) \; & \mbox{if $\;
          \; \;Q_{cr}^\prime\leq Q\leq 1\!-\!r   $} \;\,  \quad\\
  C^{\prime}(1-Q) \;\,\, & \mbox{if $\;\;
               1\!-\!r \leq Q \leq 1 $} \quad\, 
 \end{array}
\right.
\end{eqnarray}
where  $ P_{II}(Q)$, $Q_{cr}^{\prime}$ and $P_I^{\prime}(Q)$ are determined by Eqs.  (\ref{mirror1a}) - (\ref{mirror3}). 
\begin{figure}[t!]
\center{\includegraphics[width=7.5 cm]{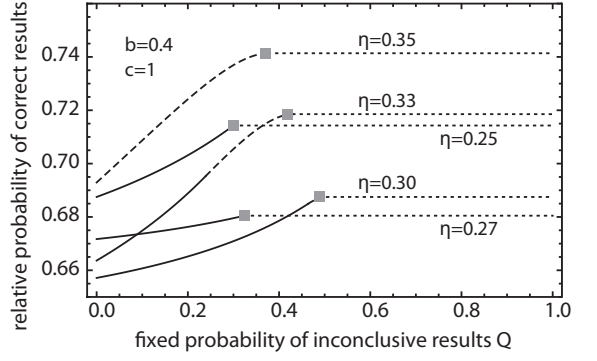} }
 \caption {Relative probability of correct results
$\frac{P_c^{max}|_Q}{1-Q}$
versus the fixed probability $Q$ of inconclusive results in an optimal measurement for discriminating the three mirror-symmetric  pure states given by Eq. (\ref{mirror}) with  $b=0.4$ for different values of the  prior probability $\eta$.}
\end{figure}

In the regions of $Q$ where the solution is given by $P_{II}(Q)$, corresponding to the dashed lines in Fig. 3,  only the states $|\psi_1\rangle$ and  $|\psi_2\rangle$ are guessed to occur in the optimal measurement.
On the other hand, all three states are guessed if $0\leq Q< Q_{cr}$ for $C\geq C^{\prime},$ and if $0\leq Q< 1-r$ for $C\leq C^{\prime}$,
 which corresponds to the full lines in Fig. 3.
For $b=0.4$ we find that  $C\leq C^{\prime}$  if $\eta \leq \eta_0\approx 0.28.$
As outlined in the Appendix, for $b>0.1$  the first line of  Eq. (\ref{pure3prime}) does not apply, and the second line is valid in the whole range $0\leq Q\leq 1-r$ since  $Q_{cr}^{\prime}<0.$   
 The dotted lines in Fig. 3 correspond to  the limiting case of large $Q$, and the squares indicate the points where this limiting case is reached.   

For minimum-error discrimination, where  $Q=0$, the maximum probability of correct  results does not depend on the relation between  $C$ and  $C^{\prime}$. Both Eq. (\ref{solQ1}) and (\ref{pure3prime}) yield   
\begin{equation}
\label{pure30}
P_c^{max}\big|_{Q=0} \!=\!\left \{
\begin{array}{ll}\!\! (1-2\eta)\left[\frac{1-\eta(1+2 b)}{1-\eta(2+ b)}\right] \; & \mbox{if $\; 
          \eta\leq \eta_{cr}   $}  \quad\\
 \!\!\eta\left(1+2\sqrt{b(1-b}\right) \; & \mbox{if $\;
              \eta \geq \eta_{cr} $},\quad\\
 \end{array}
\right.
\end{equation}
which coincides with the result obtained  already  in  Ref. \cite{andersson}. When  $\eta < \eta_{cr}$  all three states are guessed in the measurement performing minimum-error discrimination,  while otherwise only  the states $|\psi_1\rangle$ and  $|\psi_2\rangle$ are guessed to occur.

\section{Discussion and conclusions}

Before concluding the paper, we briefly discuss the relation of our method to previous investigations of the minimum-error discrimination of  arbitrary qubit states, where $Q=0$. 
From Eq. (\ref{perp}) we obtain the representation  
\begin{equation}
\label {MX}
Z= q_j |{\pi}_j^{\perp}\rangle\langle{\pi}_j^{\perp}|+\eta_j\rho_j \quad {\rm with}\;\;q_j=
{\rm Tr \,}Z-\eta_j
\end{equation}
$(j=1,\ldots.N)$, where  due to Eq. (\ref{compl1}) for  minimum-error discrimination the condition $\sum_{j=1}^N \alpha_j
|{\pi}_j^{\perp}\rangle\langle{\pi}_j^{\perp}|
=I$ has to be satisfied with non-negative values of $\alpha_j$. Upon eliminating the operator $Z$, we obtain  from the first equality in Eq. (\ref{MX}) for any pair of states the equation 
\begin{equation}
\label {MY}
 q_k  |{\pi}_k^{\perp}\rangle\langle{\pi}_k^{\perp}| - q_j  |{\pi}_j^{\perp}\rangle\langle{\pi}_j^{\perp}| =  \eta_j\rho_j    - \eta_k \rho_k 
\end{equation}
($k,j=1,\ldots N $).
Eq. (\ref{MY}) has been recently derived in an alternative way \cite{kimura, bae1}, and  has been applied to study  the minimum-error discrimination of  qubit states using a geometric formulation  \cite{kimura, jafarizadeh,bae1,ha}. 
In contrast to this, in our method, which  refers to the general case $Q\geq 0$, the operator $Z$ is not eliminated.  Rather, our approach  essentially rests on determining  $Z$ and is therefore for $Q=0$ related to the treatments of minimum-error discrimination in Refs. \cite{hunter},\cite{samsonov} and \cite{deconinck}. 

In this paper we investigated the discrimination of $N$ mixed quantum states by an optimal measurement that yields  the maximum probability of correct  results, $P_c^{max}|_Q$, while the probability of inconclusive results is fixed at a given value $Q\geq 0$. 
 For the discrimination of qudit states in a $d$-dimensional Hilbert space,  we discussed the general properties of the optimal  measurement. Moreover, we derived the  analytical solution for  optimally discriminating with fixed $Q$ between a uniformly mixed and a pure qudit state.
In the main part of the paper we specialized on the optimal discrimination of  qubit states in a two-dimensional Hilbert space  
and developed a general method to obtain the solution. We studied the special case where the prior probabilities of the  qubit states  are equal, and we also treated the  discrimination between four or less arbitrary qubit states with fixed $Q$.  As an illustrative application of our method, we derived  explicit analytical results for  discriminating qubit states which posses a  partial symmetry.

We emphasize that apart from determining $P_c^{max}|_Q$, our method also allows to consider the various possible realizations of the optimal measurement for a given discrimination problem. 
In particular,  we found that for discriminating $N$ symmetric qubit states the maximum probability of correct results with fixed $Q$ can be for instance also achieved by a measurement where only three of the states are guessed to occur when $N$ is odd, and only two of the states when $N$ is even, instead of guessing all $N$ states.

Note added: After submitting this work a related paper \cite {nakahira2} appeared.

\appendix*
\section{Analytical solution for partially symmetric qubit states}

In this Appendix  we provide the detailed derivations for the results presented in Sec. III. 
We start by treating  the case where only states  from the first group are guessed, that is $A_2=0.$   Using   Eq. (\ref{first}) and  the  first line of  Eq. (\ref{two1a}), we obtain 
\begin{equation}
\label {z0}
\frac{z_{00}}{\eta}=s+p\sqrt{\frac{ b(1\!-\!b)}{F}},\quad \frac{z_{11}}{\eta}=1\!-\!s+p{\sqrt{F b(1\!-\!b)}}
\end{equation}
with $F=f_1/f_0$ and $s=p b+\frac{1}{2}(1-p)$.
From Eq. (\ref{z0}) it  follows  that the condition  
$\frac{z_{00}}{z_{11}}\leq \frac{r}{1-r}$  is equivalent to $F \geq F_{u},$ 
where   
\begin{equation}
\label {solQ2a}
{F_{u}} = \left[\sqrt{\frac{(r-s)^2}{4r^2 p^2 b(1-b)}+\frac{1\!-\!r}{r}}- \frac{r-s}{2r p \sqrt{b(1-b)}}\right]^{2}
\end{equation}
with $F_{u}= \frac{b(1-r)^2}{(1-b)r^2}$ for  $p=1$. 
Making use of  Eq. (\ref{f}) 
we therefore arrive at  
\begin{equation}
\label{solQ2}
F =\frac{f_1}{f_0}\!=\!\left \{
\begin{array}{ll}\!\!  1-\frac {Q}{r}>F_u  & \mbox{if $\;
        \frac{z_{00}}{z_{11}}< \frac{r}{1-r}
$}  \quad\\
 \!\!\left(1\!-\!\frac{Q}{1-r}\right)^{-1}\!\!<F_u \;\; & \mbox{if $\;
             \frac{z_{00}}{z_{11}}> \frac{r}{1-r},
  $}\quad
 \end{array}
\right.
\end{equation}
and due to Eq. (\ref{first}) we obtain the explicit result 
\begin{equation}
\label{A1}
  A_1=f_0+f_1=\left \{
\begin{array}{ll}  2-\frac{Q}{r} & \mbox{for  $\;
          F_{u} < 1   $}  \quad\\
 2-\frac{Q}{1-r}  & \mbox{for $\;
           F_{u} > 1.  $}\quad
 \end{array}
\right.
\end{equation}
In the upper line we took into account that   $ \frac{z_{00}}{z_{11}}< \frac{r}{1-r}$ corresponds to  $F_u<F=1-\frac{Q}{r}$ which implies that   $F_u <1$,  and we used  a similar relation  for the lower line. 
Since $Q\geq 0$,  Eq. (\ref{solQ2}) can only be fulfilled  and hence Eq. (\ref{z0})  can only determine the optimal solution  if 
\begin{equation}
\label{solQ3}
Q<  Q_u \;\;{\rm with}\;\;Q_u=\left \{
\begin{array}{ll}  r(1-F_{u}) & \mbox{for  $\;
          F_{u} < 1   $}  \quad\\
 (1-r)\left(1- {F_u}^{-1}\right) \; & \mbox{for $\;
           F_{u} > 1,  $}\quad
 \end{array}
\right.
\end{equation}
where $Q=Q_u$ corresponds to  $  \frac{z_{00}}{z_{11}}= \frac{r}{1-r}.$  
 From  Eqs. (\ref{z0}) and  (\ref{sol}) we obtain  the maximum probability of correct results $P_c^{max}|_Q =P_{II}(Q)$  
with 
\begin{equation}
\label{A-2}
 \frac{P_{II}}{\eta}=\left \{
\begin{array}{ll}
 \!\!1-\frac{s}{r}Q+2p\sqrt{b(1-b)}\sqrt{1-\frac{Q}{r}} \;\; & \mbox{for $\;
               F_u < 1 $}\;\;\, \quad\\
\!\! 1-\frac{1\!-\!s}{1\!-\!r}Q+2p\sqrt{b(1-b)}\sqrt{1-\frac{Q}{1\!-\!r}} \;\; & \mbox{for $\;   F_u> 1. $}\;\;\, \quad
 \end{array}
\right.
\end{equation}
 Since the expressions in the upper and lower lines of Eqs. (\ref{solQ2}) - (\ref{A-2}) are identical for $F_u=1$ and $Q=Q_u=0,$ we can replace  the conditions $F_u<1$ by  $F_u\leq 1$, and similarly  $F_u>1$ by $F_u\geq1$, if we extend  the restriction $Q<Q_u$ to $Q\leq Q_u$.
 Eq. (\ref{A-2})  describes the optimal solution provided that  $z_{00}$ and $z_{11},$ determined by   Eq. (\ref{z0}), satisfy  the positivity constraints  given by Eq. (\ref{constr1}). 
 
Analogous results can be obtained for the case where only states from the second group are guessed to occur, that is where $A_1=0$.     
We still need to study the case where states from both groups are guessed in the optimal measurement, that is where $A_1>0$ and $A_2>0$. We shall do this in the following, where we derive  the complete solutions  for the problems  discussed  in Secs. III C and III D.

{\it  Equiprobable states with equal purity.} 
In accordance with Eq. (\ref{r1}) we assume that $r\geq \frac{1}{2}$  and $c>b$, which for $\eta=\eta^{\prime}=1/N$ and $p=p^{\prime}$ corresponds to $ C \geq C^{\prime}$.
Supposing that states from both groups are guessed, that is $A_1,A_2>0$, 
we obtain from Eq. (\ref{two1a})  the solution  ${z}_{00}= {z}_{11}=\frac{1+p}{2N}$ which clearly satisfies the positivity  constraints  given by Eq. (\ref{constr1}).   Taking into account that  $\frac{{z}_{00}}{{z}_{11}}=1\leq \frac{r}{1-r}$,  
Eqs. (\ref{compl2a}) -  (\ref{f}) yield  the  constants
$A_1=
\frac{2c-1+\frac{Q}{r}(1-c)}{c-b}$
 and 
$A_2= \frac{1-2b-\frac{Q}{r}(1-b)}{c-b}$.
Since  $c\geq \frac{1}{2}$, cf. Eq. (\ref{r1}),  $A_1$ cannot be negative. On the other hand,  the condition 
$A_2\geq 0$ requires that $Q$ does not exceed the critical value $Q_{cr}= r \frac{1-2b}{1-b}$.  
With the help of  Eq. (\ref {sol}) we thus arrive at the first line of  Eq. (\ref{solQ}). 
The second line of  Eq. (\ref{solQ}) refers to the case where the solution determined by Eqs. (\ref{z0}) and (\ref{A-2}) is optimal, as will be shown below by verifying that Eq.  (\ref{constr1}) is satisfied. The third line corresponds to   Eq.  (\ref{lim2}).   
With the help of  Eq. (\ref{solQ2a}) we find after a little algebra that  $F_u\leq \frac{b}{1-b}$ if $r\geq \frac{1}{2}$ and $c>b$. 
Hence for  $b<\frac{1}{2}$   it follows that $F_u<1,$ which means that for $Q_{cr}>0$ always the first line  of Eq. (\ref{A-2})
applies. 

It remains to be shown  that for $Q_{cr}\leq Q \leq Q_u$ 
  the positivity constraints given  by Eq.  (\ref{constr1}) are fulfilled when $z_{00}$ and $z_{11}$ are determined by Eq. (\ref{z0}).   
From  Eqs.   (\ref{constr2}) and  (\ref{constr2a}) with $\eta=\eta^{\prime}$ and $p=p^{\prime}$ we find after minor algebra that Eq.  (\ref{constr1}) is  satisfied
 if   $F \leq \frac{b}{1-b}$, which  because of Eq. (\ref{solQ2}) yields the two conditions
 $F=1-\frac {Q}{r}\leq  \frac{b}{1\!-\!b}$ if
$F_u\leq 1$ and
$F= ( 1\!-\!\frac{Q}{1-r})^{-1}\leq \frac{b}{1\!-\!b}$ if
            $F_u \geq 1.$
The first condition requires that $Q\geq Q_{cr}$. The second condition is fulfilled for  $ Q\leq Q_u$, as becomes obvious from  the second line of Eq. (\ref{solQ2}) and from the validitiy of  the relation  $F_u\leq \frac{b}{1-b}.$  Hence  Eq.   (\ref{constr1}) is indeed satisfied and we have derived Eq. (\ref{solQ}).

{\it  Pure states with different prior probabilities.} 
For  $p=p^{\prime}=1$ we obtain with the help of Eq. (\ref{maxconf1}) the relation
\begin{equation}
\label {pure4a}
 C-  C^{\prime} = \frac{r(\eta-\eta^{\prime})+ (2r-1)(\eta^{\prime} c-\eta b)}{r(1-r)}=\frac{d_0}{1-r}-\frac{d_1}{r},
\end{equation}
where $d_0=\eta(1\!-\!b)-\eta^{\prime}(1\!-\!c)$ and $d_1=\eta^{\prime} c-\eta b$,  while $r=N_1\eta b + N_2 \eta^{\prime}c$
with $N_1\eta+N_2\eta^{\prime}=1$.
Supposing that states from both groups are guessed,  we get  from
Eq.   (\ref{two1a})  the solutions
 ${z}_{00}= \eta\eta^{\prime}\frac{c-b}{d_0}$ and ${z}_{11}= \eta\eta^{\prime}\frac{c-b}{d_1}$ which have to positive when  Eq.  (\ref{constr2}) holds.
To be specific, we again assume that $c>b$. 
 The condition
 $ C\geq   C^{\prime}$ then implies that  $ \frac{d_1}{d_0}= \frac{ z_{00}}{ z_{11}}\leq \frac{r}{1-r}$, 
and a corresponding relation is valid when  $ C\leq   C^{\prime}$. From  Eq. (\ref{sol}) we obtain the solution    
\begin{equation}
\label {Pcmax2}
 P_c^{max}\big|_Q\!= \!\left \{
\begin{array}{ll}
\!\! P_I (Q)\!= \eta\eta^{\prime} \frac{c-b}{d_0}\left(\frac{d_0}{d_1}+1-\frac{Q}{r}\right) \; \;\,\;\;\mbox{for $
           C\geq   C^{\prime} $}\\
\!\! P_I^{\prime} (Q)\!= \eta\eta^{\prime} \frac{c-b}{d_1}\left(\frac{d_1}{d_0} +1-\frac{Q}{1-r}\right) \, \;
         \mbox{for $
           C\leq   C^{\prime}, $}
 \end{array}
\right.\quad
\end{equation}
 which  holds true for certain regions  of $Q$  where  the positivity constraint given by Eq. (\ref{constr1}) is satisfied and where  Eqs. (\ref{compl2a}) - (\ref{f}),  resulting from the completeness relation, can be fulfilled  with positive values of $A_1$ and $A_2$. Outside these regions only states from one of the groups will be guessed in the optimal measurement.  

Now we specialize on the discrimination of  three  mirror-symmetric pure states, given by  Eq. (\ref{mirror}),  where $\eta^{\prime}=1-2\eta$ and  $r= 2 \eta b +\eta^{\prime}=1-2\eta (1-b)$.    
Eq. (\ref{maxconf1}) yields the  respective maximum confidences  $C=\eta\frac{r+b(1-2r)}{r(1-r)}$ for $|\psi_1\rangle$ and $|\psi_2\rangle$,  and  $C^{\prime}=\frac{1-2\eta}{r}$ for the state $|\psi_3\rangle$. 
Provided  that all three states are guessed, that is $A_1,A_2 >0$, we obtain from Eq. (\ref{two1a}) the solution  
${z}_{00} = \eta^{\prime}$
and  ${z}_{11}=\frac{\eta^{\prime}\eta(1-b)}{1-\eta(2+b)}$. 

For   $C\geq C^{\prime}$, that is for   $\eta \geq \eta_0=\frac{1}{2+4b}$,   it follows that   the relation $\frac{{z}_{00}}{\bar{z}_{11}}\leq \frac{r}{1-r}$ holds true when all three states are guessed. 
Eqs. (\ref{compl2a}) - (\ref{f})  then yield the solutions   $ A_1= \frac{{\rm Tr}Z -\eta}{z_{00}-\eta b}$ and  $A_2= \frac{1}{r}(Q_{cr}-Q)$. Here we introduced a critical value  $Q_{cr}$, given by      
\begin{equation}
\label{crit}
\frac{Q_{cr}}{r} = 1 - \frac{\eta^2b(1- b)}{[1-\eta(2+b)]^2}, 
\quad{\rm where} \;\; Q_{cr}\geq0\;\; {\rm if}\;\;\eta \leq \eta_{cr}
\end{equation}
with  $ \eta_{cr}=\left[2+b+\sqrt{b(1-b)}\right]^{-1}$. While $A_1$ is always positive, the condition $A_2\geq 0$ requires that  $Q \leq  Q_{cr}.$  Due to Eq. (\ref {sol}) we find that  $P_c^{max}|_Q= P_I(Q)$ for $0 \leq  Q \leq Q_{cr},$ where
\begin{equation}
\label{mirror0}
P_I(Q)= (1-2\eta)\left[\frac{1-\eta(1+2 b)}{1-\eta(2+ b)}-\frac{Q}{r}\right]. 
\end{equation}
Clearly, this result only applies when $Q_{cr}\geq0,$ or  $\eta\leq\eta_{cr}$, respectively, which implies that $\eta(2+b)<1$ and that Eq. 
(\ref{constr1}) is therefore satisfied. 
As will be shown below, for $ Q_{cr}\leq Q \leq Q_u$  a measurement is optimal where $A_2=0$,  that is where  Eq. (\ref{z0}) determines the solution. Extending  Eq.  (\ref{A-2}) to the case $F_u=1$,  we obtain  $P_c^{max}|_Q=  P_{II}(Q)$ for $Q_{cr}\leq Q\leq Q_u$, where
\begin{equation}
\label{mirror1a}
 \frac{P_{II}}{\eta}=\left \{
\begin{array}{ll}
 \!\!1-\frac{b}{r}Q+2\sqrt{b(1-b)}\sqrt{1-\frac{Q}{r}} \;\; & \mbox{if $\;
                \eta\leq\eta_{cr} $}\;\;\, \quad\\
\!\! 1-\frac{1\!-\!b}{1\!-\!r}Q+2\sqrt{b(1-b)}\sqrt{1-\frac{Q}{1\!-\!r}} \;\; & \mbox{if $\;   \eta \geq \eta_{cr}.   $}\;\;\, \quad
 \end{array}
\right.
\end{equation}
Hence  we have obtained the first two lines of  Eq. (\ref{solQ1}), and the third line follows again from Eq. (\ref{lim2}). 
In Eq. (\ref{mirror1a}) we took into account that with  $r=1-2\eta(1-b)$ the condition $F_u =  \frac{b(1-r)^2}{(1-b)r^2} \leq 1 $ is equivalent to the condition $\eta \leq \eta_{cr}$.   
We  note that for  $\eta \geq\eta_0$  the condition $ \eta \leq\eta_{cr}$, or  $Q_{cr}\geq 0$,  respectively,  can be only fulfilled if   $\eta_0 \leq \eta_{cr}$, which requires that  $b\geq 0.1.$ For $b< 0.1$  we therefore  get  $\eta  >\eta_{cr}$,  or $Q_{cr}< 0$, respectively, which means that the solution is given by the lower line of Eq. (\ref{mirror1a}) in the whole range $0\leq Q\leq Q_u.$ 

Next we treat the case where $C\leq C^{\prime}$, or  $\eta \leq \eta_0$, respectively, which means that  $\frac{{z}_{00}}{{z}_{11}}\geq \frac{r}{1-r}$ if all three states are guessed. From   Eqs. (\ref{compl2a}) - (\ref{f})  we  obtain    $ A_1= (1-\frac{Q}{1-r})\frac{{\rm Tr}Z -\eta}{z_{00}-\eta b}$ and  $A_2=\frac{Q_{cr}}{r}+\frac{Q}{1-r} \frac{r-Q_{cr}}{r}$. 
It is useful to introduce $Q_{cr}^{\prime}$ with $\frac{Q_{cr}^{\prime}}{1-r}=-\frac{ Q_{cr}  }{r- Q_{cr}}$, that is   
\begin{equation}
\label{pure40}
\frac{Q_{cr}^{\prime}}{1-r}= 1 - \frac{[1-\eta(2+b)]^2}{\eta^2b(1- b)},  \quad{\rm where} \;\; Q_{cr}^{\prime}\geq 0\;\; {\rm if}\;\;\eta \geq \eta_{cr}.   
\end{equation}
Provided that $Q_{cr}^{\prime}\leq Q\leq 1-r$ the conditions  $A_1\geq 0$ and $A_2 \geq 0$ hold true.  
Using  Eq. (\ref {sol}) we therefore get   $P_c^{max}|_Q=P_{I}^{\prime}(Q)$ for $Q_{cr}^{\prime} \leq Q\leq 1-r$, where
\begin{equation}
\label{mirror3}
P_{I}^{\prime} (Q) =\frac{1-2\eta}{1-\eta(2+ b)}\left[1-\eta(1+2 b)-\eta(1-b)\frac{Q}{1-r}\right].
\end{equation}
Thus we have obtained the second line of  Eq. (\ref{pure3prime}). 
For $Q=1\!-\!r$, where  $A_1=0$, only the state $|\psi_3\rangle$ is guessed. From Eq. (\ref{mirror3}) it follows that $P_{I}^{\prime} (Q)=C^{\prime}(1-Q)$ when $Q=1\!-\!r.$ This means that  in Eq. (\ref{lim2})  $Q_u$ corresponds to $1\!-\!r$, and we arrive at the third line of Eq. (\ref{pure3prime}).  On the other hand, for $Q=Q_{cr}^{\prime}$ we get $A_2=0.$
 As will be shown below,  when  $Q\leq Q_{cr}^{\prime}$  a measurement is optimal where  $A_2=0$ and where the solution is therefore  determined  by  $P_{II}(Q)$,  given by  Eq. (\ref{mirror1a}). This yields the first line of   Eq. (\ref{pure3prime}). 
 For   $\eta\leq\eta_0$  the condition  $ \eta \geq \eta_{cr}$, or $Q_{cr}^{\prime}\geq 0$, respectively,  can be only fulfilled if   $\eta_{cr} \leq \eta_{0}$, which requires that  $b\leq 0.1.$ For $b> 0.1$  we therefore  get $Q_{cr}^{\prime}< 0$,  which means that the solution is given by Eq. (\ref{mirror3}) in the whole range $0 \leq Q\leq 1-r$.

 Let us finally consider  the positivity constraints given  by Eq.  (\ref{constr1}) for the case where the solution is  described  by $P_{II}(Q)$, that is where  $z_{00}$ and $z_{11}$ are determined by Eq. (\ref{z0}).  Making use of Eqs.  (\ref{constr2}) and (\ref{constr2a}) with $p=1$ and $c=1$ we find after minor algebra that Eq.  (\ref{constr1}) is satisfied  for 
$F \leq \frac{\eta^2b(1\!-\!b)}{[1-\eta(2+b)]^2 }.$   Taking the definition of $Q_{cr}$  into account, we therefore obtain from  Eq. (\ref{solQ2}) the two conditions  
 $F=1-\frac {Q}{r}<1- \frac {Q_{cr}}{r}$ if
$F_u\leq 1$ and
$F=({1\!-\!\frac{Q}{1-r}})^{-1}< 1- \frac {Q_{cr}}{r} $ if
            $F_u\geq 1.$
For   $C\geq C^{\prime}$, or  $\eta\geq\eta_0$, respectively, the first condition requires that $Q\geq Q_{cr}$, while the second condition is always fulfilled for $Q\leq Q_u $, as follows from  the second line of  Eq. (\ref{solQ2}) and from the relation  $F_u \leq 1- \frac {Q_{cr}}{r}$  which holds for $\eta\geq\eta_0$.   For  $C\leq C^{\prime}$  the second condition requires that $Q \leq Q_{cr}^{\prime}$ since $1- \frac {Q_{cr}}{r}=({1\!-\!\frac{Q_{cr}^{\prime}}{1-r}})^{-1}$, which means that it only applies if $Q_{cr}^{\prime}>0$, or $Q_{cr}<0$, respectively, where the first condition is always satisfied.  
 Hence Eq.  (\ref{constr1}) indeed restricts the range of validity of Eq. (\ref{z0}) to the regions  $0\leq Q \leq Q_{cr}^{\prime}$ if $C\leq C^{\prime}$ and $Q_{cr}\leq Q\leq Q_u$ if $C\geq C^{\prime}.$

\section*{Acknowledgments}
The author acknowledges partial financial support by Deutsche Forschungsgemeinschaft DFG (SFB 787).

\end{document}